\documentclass[english,aps,prl,amsmath,twocolumn,superscriptaddress,showpacs,notitlepage]{revtex4-2}
\usepackage[T1]{fontenc}
\usepackage[latin9]{inputenc}
\usepackage{epsfig}
\usepackage{subfig}

\DeclareMathAlphabet\mathbfcal{OMS}{cmsy}{b}{n}

\makeatletter

\usepackage{mathptmx,newtxtext,newtxmath,xspace}
\usepackage{amsbsy,bm,bbold}
\usepackage{graphicx,color,xcolor,epsfig,rotate}
\usepackage{fancyhdr}
\usepackage[colorlinks=true, 
            linkcolor=blue, 
            urlcolor=blue,
            citecolor=blue]{hyperref}
\usepackage{soul}
\usepackage{cancel}
\newcommand{\iu}{{i\mkern1mu}}

\makeatother

\usepackage{babel}
\usepackage{ragged2e}

\begin{document}
\title{ Skyrmions of Frustrated Quantum Dimer Systems}
\author{Fletcher~Williams}
\email{fwilli18@vols.utk.edu}
\affiliation{Department of Physics and Astronomy, The University of Tennessee, Knoxville, Tennessee 37996, USA}
\author{David~Dahlbom}
\affiliation{Department of Physics and Astronomy, The University of Tennessee, Knoxville, Tennessee 37996, USA}
\affiliation{Spallation Neutron Source, Oak Ridge National Laboratory, Oak Ridge, Tennessee 37830, USA}
\author{Hao~Zhang}
\affiliation{Department of Physics and Astronomy, The University of Tennessee, Knoxville, Tennessee 37996, USA}
\affiliation{Theoretical Division and CNLS, Los Alamos National Laboratory, Los Alamos, New Mexico 87545, USA}
\author{Shruti~Agarwal}
\affiliation{Department of Physics and Astronomy, The University of Tennessee,
Knoxville, Tennessee 37996, USA}
\author{Kipton~Barros}
\affiliation{Theoretical Division and CNLS, Los Alamos National Laboratory, Los Alamos, New Mexico 87545, USA}
\author{Cristian~D.~Batista}
\email{cbatist2@utk.edu}
\affiliation{Department of Physics and Astronomy, The University of Tennessee,
Knoxville, Tennessee 37996, USA}
\affiliation{Quantum Condensed Matter Division and Shull-Wollan Center, Oak Ridge
National Laboratory, Oak Ridge, Tennessee 37831, USA}
\date{\today}
\begin{abstract}
Magnetic skyrmions are topologically protected solitons observed in various classes of real magnets. In two-dimensional systems, where the target space of local magnetization values is the two-sphere $S^2$, skyrmion textures are classified by the homotopy classes of two-loops  $S^2$ in $S^2$: $\Pi_2(S^2) \cong Z$. Here, we demonstrate that more general topological skyrmion textures emerge in the classical limit of quantum dimer systems, where the phase space of the relevant classical theory is  $\mathbb{CP}^{N-1}$ (with $N=4$ for the case of interest), because the relevant second homotopy group, $\Pi_2(\mathbb{CP}^{N-1}) \cong Z$ for $N\geq 2$,  remains unchanged. Building on the framework established by Zhang et al. (2023), we consider a classical limit based on SU(4) coherent states, which preserve intra-dimer entanglement. We show that the zero-temperature phase diagram of frustrated spin-dimer systems on a bilayer triangular lattice with weak inter-dimer coupling includes two magnetic-field-induced $\mathbb{CP}^{3}$ skyrmion crystal phases.  
\end{abstract}
\maketitle

\section{Introduction}
Since their experimental discovery in 2009 ~\cite{Muhlbauer2009} and in several other chiral magnets in the following years ~\cite{Yu2010a,Yu2011,Seki2012,Adams2012}, magnetic skyrmions have excited the condensed matter community as an area rich both theoretically,  in deepening our understanding of the important role of topology in physics, and practically,  in their promise for spintronics data storage and logic technology ~\cite{Romming2013}.

The skyrmions that have been predicted ~\cite{Bogolubskaya89, Bogolyubskaya89b, Bogdanov1994, Okubo12, Leonov2015, Lin2016_skyrmion, Hayami16, Batista16, Wang2020_RKKY, Hayami2021_review} and observed~\cite{Muhlbauer2009, Yu2010a, Yu2011, Seki2012, Adams2012, Yu2012, Yu2014_biskyrmion, Mallik1998_paramana, Saha1999, Kurumaji2019, Chandragiri_2016, Hirschberger2019} in magnetic materials can be classified using the second homotopy group of the sphere $S^2 \approx\;\mathbb{CP}^1$: $\Pi_2(S^2)=Z$. Homotopy classes of mappings, $F: S^2 \to S^2$, from the sphere to the sphere,  arise from considering 2D magnetic textures which can be regarded as mappings, $\boldsymbol{m}(\boldsymbol{r})$, from the 2D plane $\mathbb{R}^2$ (base manifold) to points on the Bloch sphere representing SU(2) spin coherent states. If we only consider magnetization textures $\boldsymbol{m}(\boldsymbol{r})$ that are constant for $|\boldsymbol{r}| \to \infty$, we can map all the points $|\boldsymbol{r}| \to \infty$ to a single point and note that the 2D plane plus a point at infinity  is topologically equivalent to the two-sphere: $\mathbb{R}^2 \cup\{\infty\} \approx S^2$. Topologically distinct smooth mappings can then be labeled by integers that represent the number of times the mapping is ``wrapping'' the sphere.

Building on the analytical framework for taking classical limits of spin systems using coherent states of completely symmetric irreps of SU($N$)~\cite{Hao21}, Zhang et al.~\cite{Hao23} showed how exotic  $\mathbb{CP}^2$ skyrmions could be realized in realistic quantum magnets using coherent states of SU(3). In general, since the manifold of  coherent states of completely symmetric irreps of SU($N$) is $\mathbb{CP}^{N-1}$ and  $\Pi_2(\mathbb{CP}^{N-1})=\mathbb{Z}, N\geq2$, we can in principle find different types of skyrmions in magnetic systems~\cite{Hao23,akagi2021}. The basic challenge is to find realistic spin Hamiltonians that can naturally lead to the emergence of these exotic topological textures.

Related generalized skyrmion crystals have also been proposed in
non-Kramers doublet systems, where the relevant local degrees of
freedom are multipolar rather than conventional magnetic
dipoles~\cite{Zhang2024}. In particular, these textures involve
spatially varying quadrupolar and octupolar moments. The present
$\mathbb{CP}^{3}$ skyrmions provide a distinct realization in which
the enlarged local manifold arises from the complete singlet-triplet
Hilbert space of an entangled spin-$1/2$ dimer.

Here, we present the first example of a model that produces skyrmions at zero temperature solely through frustrated but isotropic Heisenberg exchange interactions and an applied magnetic field. Spin anisotropy, in the form of  Dzyaloshinskii-Moriya interactions or 
easy-axis single-ion or exchange anisotropy, has generally been considered necessary for skyrmion stabilization at zero temperature~\cite{Bogdanov1994, Leonov2015, Batista16}. To the best of our knowledge, the seminal work by Okubo et al.~\cite{Okubo12} is the only prior study to report the emergence of field-induced $\mathbb{CP}^1$ skyrmions \emph{at finite temperature} in a frustrated isotropic Heisenberg model.

Furthermore, the target manifold of these new skyrmions is $\mathbb{CP}^{N-1}$, which is \textit{not} the order parameter space of a magnetically ordered state, as it is for the usual $\mathbb{CP}^1$ case. In the particular $N=4$ case we explore,  $\mathbb{CP}^{3}$ is the manifold of single-dimer quantum mechanical states spanned by the $S=1$ triplet states and the singlet state, which breaks no symmetry at all.
This observation implies that the core or periphery region of  $\mathbb{CP}^{N-1}$ skyrmions can be completely isotropic or paramagnetic. In other words, the sphere (base manifold) that is embedded in $\mathbb{CP}^{N-1}$ may have a singlet state at the south or north pole. 

In the general case, $\mathbb{CP}^{N-1}$ manifolds may arise from local Hilbert spaces with more than one singlet state and both the south and the north poles may correspond to two orthogonal singlets.
While such skyrmions locally break the time-reversal symmetry (the skyrmion density is proportional to the local Berry curvature), they would remain hidden to standard dipolar probes, such as neutron diffraction. 
Nevertheless, through spin--electron or spin--phonon coupling, such
textures may induce momentum-space Berry curvature in electronic or
phononic bands. The magnitude and sign of this induced Berry curvature
depend on the microscopic coupling and the corresponding band
structure.

In this work, we consider a family of models that are isotropic and realistically attainable. 
While we focus on a triangular lattice of antiferromagnetically
coupled spin-$1/2$ dimers with competing ferromagnetic and
antiferromagnetic inter-dimer interactions, the characteristic
length scale of the skyrmion textures can be made much larger than the
lattice spacing. Their stabilization can therefore be described
by a long-wavelength continuum theory whose leading terms are
determined by symmetry rather than by the microscopic details of
a particular lattice~\cite{LinHayami2016}. We consequently expect
the mechanism identified here to extend to other hexagonal
lattices of antiferromagnetically coupled dimers, including
honeycomb and kagome lattices, provided that their exchange
spectra exhibit the required set of symmetry-related
finite-wave-vector minima and that the corresponding low-energy
pseudospin model has effective easy-axis anisotropy.

After first determining the classical limit of our model Hamiltonian, we present the numerically derived $T=0$ phase diagram, which includes two field-induced $\mathbb{CP}^3$ skyrmion crystal phases. Both of these extend beyond the perturbative regime, and one of them extends significantly outside the perturbative regime.

\section{Model}

\begin{figure}
    \centering
    
    \includegraphics[width=0.5\linewidth]{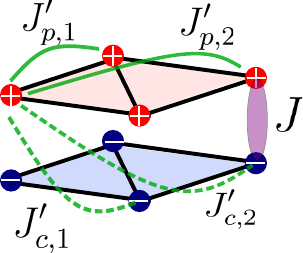}
    \captionsetup{justification=Justified}
    \caption{ {\bf Exchange interactions.} Intra-site and inter-site exchange interactions up to second nearest neighbor on a triangular lattice.}
    \label{fig:two_dimers}
\end{figure}

Consider the model of a triangular lattice (TL) composed of coupled
dimers, as illustrated in Fig.~\ref{fig:two_dimers},
where each dimer consists of two $S=1/2$ spins. The
operator $\hat{\bm S}_{j,\sigma}$ denotes a standard $S=1/2$ spin
operator, where $j$ labels a dimer of the triangular lattice and
$\sigma=+$ ($-$) labels the spin in the upper (lower) layer.
The Hamiltonian can be decomposed into terms describing intra-dimer
interactions and terms corresponding to inter-dimer interactions
between nearest-neighbor (NN) and next-nearest-neighbor (NNN) dimers:

\begin{equation}
\hat{\mathscr{H}} = \hat{\mathscr{H}}_0 + \hat{\mathscr{H}}', \nonumber
\end{equation}
where
\begin{equation}
\hat{\mathscr{H}}_0= J \sum_{j}\left(\boldsymbol{{\hat S}}_{j+} \cdot \boldsymbol{{\hat S}}_{j-}-\frac{1}{4}\right)-g \mu_B B \sum_{j, \sigma_{\pm}} {\hat S}_{j \sigma}^z,
\label{hamil}
\end{equation}
and 
\begin{eqnarray}
    \hat{\mathscr{H}}'=\sum_{\langle j,l \rangle_{\nu},\sigma=\pm}\left[\frac{J_{p\nu}^{\prime}}{2}\left(\boldsymbol{{\hat S}}_{j \sigma} \cdot \boldsymbol{{\hat S}}_{l \sigma}\right)+\frac{J_{c\nu}^{\prime}}{2}\left(\boldsymbol{{\hat S}}_{j \sigma} \cdot \boldsymbol{{\hat S}}_{l \bar{\sigma}}\right)\right]. 
\label{eq:hamil_perturbation}
\end{eqnarray}
The index $\nu=1,2$ indicates nearest-neighbor (NN) and next-nearest-neighbor (NNN) interactions, respectively, with $\langle j,l \rangle_1$ ($\langle j,l \rangle_2$) denoting pairs of NN (NNN) dimers. The Hamiltonian term ${\hat{\mathscr{H}}}_0$ includes an antiferromagnetic (AFM) intra-dimer exchange $J>0$ and a Zeeman coupling to an external magnetic field. We set $|J|$ as the unit of energy, i.e., $J = 1$. The perturbation term ${\hat {\mathscr{H}}}'$ introduces couplings between dimers, with distinct exchange constants for parallel and crossed interactions. For Mott insulators, the interaction strength between dimers is expected to decay exponentially with the inter-dimer distance. Accordingly, we retain only interactions up to next-nearest neighbors and neglect longer-range couplings. The NN (NNN) exchange interactions are assumed to be ferromagnetic (antiferromagnetic), namely $J_{p1}', J_{c1}' < 0$ and $J_{p2}', J_{c2}' > 0$. The subscripts $p$ and $c$ refer to parallel and crossed inter-dimer interactions, as illustrated in Fig.~\ref{fig:two_dimers}. In view of the symmetry of ${\hat{\mathscr{H}}}_0$ under a global exchange of the two spins in each dimer ($\boldsymbol{{\hat S}}_{j \sigma} \to \boldsymbol{{\hat S}}_{j \bar{\sigma}}$), it will also be convenient to define the parameter,
\begin{equation}
 J^{\pm}_{\nu} :=  \frac{J_{p \nu}^\prime \pm J_{c\nu}^\prime}{2}.
\end{equation}

We proceed to take the classical limit of this Hamiltonian using a generalized classical limit for quantum spin systems ~\cite{Perelomov1972,Hao21, David2022, David2022b, David2023} that preserves the entanglement between the sites of each dimer. We begin by assuming that the many-body quantum state is a direct product of SU(4) coherent states ~\cite{Gnutzmann98, Papanicolaou1988, Batista04, Zapf06, Lauchli2006, Muniz14, Galkina14, David2024}:
\begin{equation}
|\bm{\mathrm{Z}} \rangle=\otimes_{j}|\bm{\mathrm{Z}}_j \rangle.
\label{eq:z}
\end{equation}
SU(4) coherent states are chosen since they are the simplest states capable of completely accounting for the entanglement between two $S=1/2$ spins.  We will further express the Hamiltonian in terms of the generators of SU(4) coherent states, i.e. in terms of the local $\mathfrak{su}(4)$ Lie algebra on each dimer.  
Specifically, define generators, ${\hat T}^{\alpha}$, as follows:
\begin{eqnarray}
    {\hat T}^{1} &=& {\hat S}_+^{x} \quad \quad \quad {\hat T}^{2} = {\hat S}_+^{y} \quad \quad \ \ \ {\hat T}^{3} \, = {\hat S}_+^{z} \nonumber \\
    {\hat T}^{4} &=& {\hat S}_-^{x} \quad \quad \quad {\hat T}^{5} = {\hat S}_-^{y} \quad \quad \ \ \ {\hat T}^{6} \, = {\hat S}_-^{z} \nonumber \\
    {\hat T}^{7} &=& 2{\hat S}_+^{x} {\hat S}_-^{x} \quad \ \ {\hat T}^{8} = 2 {\hat S}_+^{y} {\hat S}_-^{y} \quad {\hat T}^{9} \ = 2 {\hat S}_+^{z} {\hat S}_-^{z} \label{eq:basis} \\
    {\hat T}^{10} &=& 2{\hat S}_+^{y} {\hat S}_-^{z} \quad {\hat T}^{11} = 2 {\hat S}_+^{z} {\hat S}_-^{y} \quad {\hat T}^{12} = 2 {\hat S}_+^{z} {\hat S}_-^{x} \nonumber \\
    {\hat T}^{13} &=& 2{\hat S}_+^{x} {\hat S}_-^{z} \quad {\hat T}^{14} = 2 {\hat S}_+^{x} {\hat S}_-^{y} \quad {\hat T}^{15} = 2 {\hat S}_+^{y} {\hat S}_-^{x}. \nonumber 
\end{eqnarray}
These generators satisfy the orthonormality condition,
\begin{equation}
{\rm Tr}{{\hat T}^{\mu}{\hat T}^{\nu}} = \delta_{\mu \nu},
\end{equation}
and are complete in the sense that any observable on a 4-dimensional Hilbert space may be written as $c_0 {\hat 1} + c_{\mu} {\hat T}^{\mu}$ for real coefficients $c_0$ and $c_{\mu}$. (Here, as throughout the paper, generator indices run from 1 to 15 and summation over repeated Greek indices is assumed.) The two terms of the Hamiltonian, Eqs~\eqref{hamil} and~\eqref{eq:hamil_perturbation}, may now be rewritten in terms of these generators as
\begin{equation}
\hat{\mathscr{H}}_0= \frac{J}{2} \sum_{j}\left({\hat T}^7_{j}+{\hat T}^8_{j}+{\hat T}^9_{j} -\frac{1}{2}\right)-g \mu_B B \sum_{j}\left({\hat T}^3_j+{\hat T}^6_j\right)
\end{equation}
and 
\begin{eqnarray}
    \hat{\mathscr{H}}^{\prime}&=&\sum_{\langle j,l \rangle_{\nu}}\left[\sum_{\mu=1}^{6}\frac{J_{p,jl}^{\prime}}{2}{\hat T}_j^\mu  {\hat T}_l^\mu +\sum_{\mu=1}^{3} \frac{J_{c, jl}^{\prime}}{2}\left({\hat T}_j^\mu  {\hat T}_l^{\mu+3} + {\hat T}_j^{\mu+3}  {\hat T}_l^\mu \right) \right]. \nonumber
\end{eqnarray}

Due to the product state assumption of Eq.~\eqref{eq:z}, the classical limit of the Hamiltonian $\hat{\mathscr{H}} $ can be expressed as a function  of the ``color field,''
\begin{equation}
n^{\mu}_j \equiv \langle \bm{\mathrm{Z}}_j| {\hat T}_{j}^{\mu} |\bm{\mathrm{Z}}_j\rangle,
\label{eq:cfield}
\end{equation}
 Equivalently, the operator representation of the color field is
\begin{equation}
\boldsymbol{\mathfrak{n}}_j
\equiv n_j^\mu \hat{T}_j^\mu
=
|\bm{\mathrm{Z}}_j\rangle\langle\bm{\mathrm{Z}}_j|
-\frac{\mathbb{1}_4}{4}.
\label{eq:color-operator}
\end{equation}
Thus, the color field $\boldsymbol{\mathfrak{n}}_j$ and the coherent
state $|\bm{\mathrm{Z}}_j\rangle$ provide equivalent representations
of the same classical state~\cite{David2022,David2022b}.

The components of the color field defined in Eq.~~\eqref{eq:cfield} satisfy the constraints
\begin{equation}
n^{\mu} n^{\mu}=\frac{3}{4}, \quad n^{\mu}= 4
d_{\mu \nu \eta} n^{\nu} n^{\eta},
\label{eq:const}
\end{equation}
where $d_{\mu \nu \eta}={\text {Tr}}\left( \hat T^\mu \{ \hat T^\nu,\hat T^\eta\} \right)/4.$ These constraints  lead to the Casimir identity,
\begin{equation}
d_{\mu\nu\eta}n^\mu n^\nu n^\eta = \frac{3}{16},
\label{eq:Casimir}
\end{equation}
whose derivation is provided in Appendix~\ref{app:SUN}. Here 
$n^\mu$ is a 15-element real vector. The first six components correspond to the expected dipoles on each site of each dimer. The remaining nine components  are defined by the expectation value of  bilinear forms related to local observables of each dimer: the singlet character associated with $1/4 -{\hat {\bm S}}_+ \cdot {\hat {\bm S}}_-$ (one component), the quadrupolar moment $Q^{\mu \nu}={\hat S}^\mu_+ {\hat S}^\nu_- + {\hat S}^\nu_+ {\hat S}^\mu_- - 2 {\hat {\bm S}}_+\cdot {\hat {\bm S}}_-/ 3$ (five components), and the spin current density ${\hat {\bm S}}_+ \times {\hat {\bm S}}_-$ (three components). 
The expectation values defining the components $n^\mu$ are evaluated
in the limit $M\to\infty$, where $M$ labels the completely symmetric
irreducible representations of SU(4). We emphasize that this is not
the conventional $S\to\infty$ limit of the microscopic spins, which
remain $S=1/2$. Formally, after rescaling the SU(4) generators by
$1/M$, their commutators and relative variances scale as $1/M$.
Consequently, the coherent-state expectation values become classical
variables, and products of operators factorize in the limit
$M\to\infty$:
\[
\langle\hat{T}^\mu\hat{T}^\nu\rangle
=
\langle\hat{T}^\mu\rangle
\langle\hat{T}^\nu\rangle.
\]

By taking this classical limit of our Hamiltonian (Eq.~\eqref{hamil} and ~\eqref{eq:hamil_perturbation}), we can express the classical Hamiltonian $H_{\text {SU(4)}}$ in terms of the color field:
\begin{eqnarray}
H_{\text {SU(4)}}&=& \frac{J}{2} \sum_{j}\left(n^7_{j} + n^8_{j} +n^9_{j} -\frac{1}{2}\right)-g \mu_B B \sum_{j}\left( n^3_j+ n^6_j\right) \label{HSU4} \\
&+& \sum_{j,l}\left[\sum_{\mu=1}^{6}\frac{J_{p,jl}^{\prime}}{2}n_j^\mu  n_l^\mu +\sum_{\mu=1}^{3} \frac{J_{c, jl}^{\prime}}{2}\left(n_j^\mu  n_l^{\mu+3} + n_j^{\mu+3}  n_l^\mu \right) \right]. 
\nonumber
\end{eqnarray}
The classical equations of motion can be derived by first starting with Heisenberg's equations of motion for the quantum Hamiltonian,
\begin{equation}
    i \hbar \frac{d {\hat T}^\mu _ j}{dt} = [{\hat T}^\mu _j , {\hat{\mathscr{H}}}] = i f_{\mu \nu \eta} \frac{\partial {\hat{\mathscr{H}}}}{\partial {\hat T}^\nu} {\hat T}^{\eta}_j,
    \label{LLQ}
\end{equation}
where $f_{\mu \nu \eta}$ are the anti-symmetric structure constants of the ${\mathfrak su}$(4) Lie algebra (see Appendix ~\ref{app:structure_constants}),
\begin{equation}
[{\hat T}^\mu _j, {\hat T}^\nu _j] = i f_{\mu \nu \eta} {\hat T}^\eta _ j.
\label{eq:commutation_relations}
\end{equation}

By taking expectation values of Eq.~\eqref{LLQ} with respect to the coherent states \eqref{eq:z}, we obtain the classical equation of motion:
\begin{equation}
    i \hbar \frac{d n^\mu _ j}{dt} = i f_{\mu \nu \eta} \frac{\partial H_{\text {SU(4)}}}{\partial n^\nu} n^{\eta}_j.
    \label{LLC}
\end{equation}
This is a generalization of the Landau-Lifshitz equation of motion, which is obtained instead through a classical limit with SU(2) coherent states~\cite{Hao21}.

To analyze the topological soliton solutions of Eq.~\eqref{HSU4} we take the continuum limit of $H_{\text {SU(4)}}$ by assuming that the characteristic wavelength of the color field is much longer than the lattice spacing $a$:
\begin{eqnarray}
    H_{\text {SU(4)}} &\approx& \int dr^2  {\cal H}_{\rm },
\end{eqnarray}
with Hamiltonian density
\begin{eqnarray}
 {\cal H}&=&\left[     {\mathcal J}_1^{\mu \nu}(\nabla n^\mu) \cdot (\nabla n^\nu) \right. 
     + \left. {\mathcal J}_2^{\mu \nu} (\nabla ^2 n^\mu)(\nabla ^2 n^\nu) + {\mathcal B}^\mu n^\mu \right]. \nonumber \\
    \label{HSU4cont}
\end{eqnarray}    
The expressions for the coupling constants ${\mathcal J}_1^{\mu \nu}$, ${\mathcal J}_2^{\mu \nu}$ are given in Appendix ~\ref{app:H_param}.
The first term of Eq.~\eqref{HSU4cont} corresponds to an anisotropic $\mathbb{CP}^3$  non-linear sigma model. The second term with four derivatives is necessary to account for the fact that the inter-dimer exchange interaction has global minima at finite wave vectors. The last term is a generalized Zeeman coupling to  an external field dictated by the magnetic field and the intra-dimer exchange $J$ that splits the singlet and triplet states of each dimer.

To analyze the skyrmion solutions of this model, we first note that the base plane $\mathbb{R}^{2}$ can be compactified to $S^2$  after assuming that the color field takes a constant value
$n_{\infty}$ at spatial infinity (boundary condition). Introducing the operator field $\boldsymbol{\mathfrak{n}} = n^\mu \hat{T}^\mu$, which allows us to eliminate the structure constants from the expression, we characterize the spin configurations by the topological charge of the mapping 
 $\boldsymbol{\mathfrak{n}}: \mathbb{R}^{2} \sim S^{2} \mapsto \mathbb{CP}^{3}$:
 
\begin{equation}
C=-\frac{i}{4 \pi} \int \mathrm{d}x \mathrm{d}y   \varepsilon_{jk}\operatorname{Tr}\left(\boldsymbol{\mathfrak{n}}\left[\partial_{j} \boldsymbol{\mathfrak{n}}, \partial_{k} \boldsymbol{\mathfrak{n}}\right]\right),
\label{eq:skxcharge}
\end{equation}
where $\epsilon_{jk}$ is the Levi-Civita symbol in two dimensions.
For the lattice systems of interest, the $\mathbb{CP}^3$ skyrmion charge can be computed after interpolating the color fields on nearest-neighbor sites $\boldsymbol{\mathfrak{n}}_j, \boldsymbol{\mathfrak{n}}_k$ and  $\boldsymbol{\mathfrak{n}}_l$ along the $\mathbb{CP}^3$ geodesic,
\begin{equation}
C=\sum_{\triangle_{jkl}}\rho_{jkl}
=
\frac{1}{2\pi}
\sum_{\triangle_{jkl}}
\left(
\gamma_{jl}+\gamma_{lk}+\gamma_{kj}
\right),
\label{eq:totsky}
\end{equation}
where $\triangle_{jkl}$ denotes each oriented triangular plaquette
of nearest-neighbor sites $j\to k\to l$, and
\[
\gamma_{kj}
=
\arg\left[
\left\langle
\bm{\mathrm{Z}}_k
\mid
\bm{\mathrm{Z}}_j
\right\rangle
\right]
\]
is the Berry connection on the bond $j\to k$~\cite{Hao23}.

In the numerical calculations, the skyrmion density on each
plaquette is evaluated as
\[
\rho_{jkl}
=
\frac{1}{2\pi}
\operatorname{Arg}\left[
\left\langle\bm{\mathrm{Z}}_j
\mid\bm{\mathrm{Z}}_l\right\rangle
\left\langle\bm{\mathrm{Z}}_l
\mid\bm{\mathrm{Z}}_k\right\rangle
\left\langle\bm{\mathrm{Z}}_k
\mid\bm{\mathrm{Z}}_j\right\rangle
\right],
\]
where $\operatorname{Arg}\in(-\pi,\pi]$ denotes the principal
branch. Equivalently, the sum of the three bond phases in
Eq.~\eqref{eq:totsky} is reduced modulo $2\pi$ to this principal
interval.

For the problem under consideration, the skyrmion density
$\rho_{jkl}$ can be interpreted as the density of the solid angle
on the Bloch sphere spanned by the singlet and $S^z=1$ triplet
states (see Appendix~\ref{app:SkDensity}).

We emphasize that the color-field formalism developed above is fully
equivalent to the coherent-state formalism. In particular, the
one-to-one correspondence given in Eq.~\eqref{eq:color-operator}
allows the topological charge to be expressed either in terms of the
operator field $\boldsymbol{\mathfrak{n}}_j$ or directly in terms of
the coherent states $|\bm{\mathrm{Z}}_j\rangle$.

The $\mathbb{CP}^{3}$ topological density is the real-space Berry
curvature associated with the complete local dimer state. Introducing
the Berry connection
\[
a_\mu(\bm{r})
=
-i\langle\bm{\mathrm{Z}}(\bm{r})
\vert\partial_\mu\bm{\mathrm{Z}}(\bm{r})\rangle,
\]
the topological charge can equivalently be written as
\[
C
=
\frac{1}{2\pi}
\int d^2r\,
\left(
\partial_x a_y-\partial_y a_x
\right).
\]
It may therefore be interpreted as a generalized
$\mathrm{SU}(4)$ scalar chirality of the complete color field,
including both dipolar and multipolar components. This quantity
should not be confused with the ordinary scalar spin chirality of
either physical layer, which probes only the dipolar projection of
the full dimer state. The familiar relation between scalar spin
chirality and real-space Berry curvature for an
$\mathrm{SU}(2)$ skyrmion is the special $$\mathbb{CP}^{1}$$ limit
of this more general geometric structure.

\section{Results}

\subsection{Phase diagram}

\begin{figure}[htp!]
\centering
\includegraphics[width=1.0\linewidth]{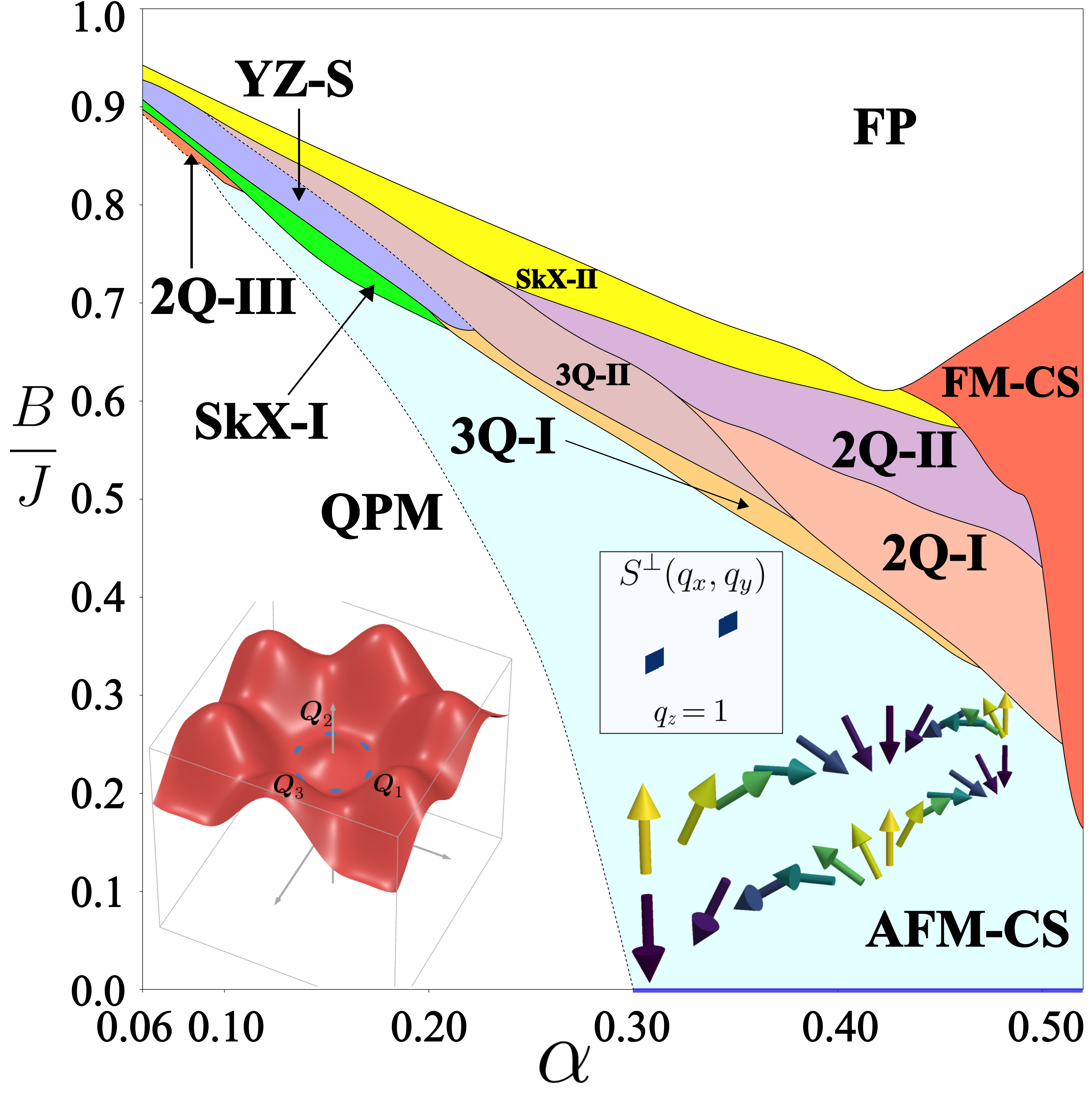}
\captionsetup{justification=Justified}
\caption{ {\bf $\bm{T=0}$ phase diagram of the classical Hamiltonian  $H_{\text {SU(4)}}$ as a function of $\alpha$ and the external field $B$}. The ordering wave vector was fixed by setting $J^{-}_2/|J^{-}_1| = 2/(1+\sqrt{5})$.  Here, $\alpha$ parametrizes the overall strength of the inter-dimer
couplings through $J_1^-=-\alpha J/2$ and
$J_2^-=\alpha J/(1+\sqrt{5})$, where
$J_\nu^-=(J'_{p,\nu}-J'_{c,\nu})/2$. Additionally, $\Delta$, the effective anisotropy of the low-energy pseudospin model, was set to $\Delta =\Delta_1 = \Delta_2=1.2$, where $\Delta>1$ corresponds to an effective easy-axis anisotropy.
The triplon dispersion is shown as an inset in the QPM phase, with the six minima indicated. The blue line for $B=0$ in the AFM-CS phase indicates a special  regime of the phase in which $SU(2)$ symmetry is preserved and thus spirals, propagating along any of the six ordering wave vector directions, can have polarization planes oriented in any direction. An example of such a spiral along with its corresponding static structure factor is provided as an inset of the AFM-CS phase. }
\label{fig:phasediagram}
\end{figure}

The zero-temperature phase diagram (Fig.~\ref{fig:phasediagram}) is obtained by numerically minimizing the classical spin Hamiltonian $H_{\text{SU(4)}}$ given in Eq.~\eqref{HSU4}. This analysis was carried out using the numerical tools provided by the open-source software package \texttt{Sunny.jl}~\cite{Dahlbom2025}. 

In Appendix~\ref{app:LowEnergyLattice}, we derive an effective low-energy pseudospin-$1/2$ model, valid in the limit of weak inter-dimer interactions, $|J'_{(p,c),\nu}|/J \ll 1$ with $\nu = 1,2$, to identify the generic conditions required for stabilizing skyrmion textures. This limit is realized by applying a sufficiently strong magnetic field $B$, which isolates two nearly degenerate low-energy singlet and $S^z = 1$ triplet states. To first order in the inter-dimer interactions, the  effective low-energy Hamiltonian $\tilde{\mathscr{H}}$ is obtained by projecting $\hat{\mathscr{H}}$ into this low-energy subspace:
\begin{eqnarray}
    \tilde{\mathscr{H}}&=& 2  \sum_{\langle j,l \rangle_{\nu}}  J^{-}_\nu \left({ \hat \tau}_j^x { \hat \tau}_{l}^x+{ \hat \tau}_j^y {\hat \tau}_{l}^y + \Delta_\nu { \hat \tau}_{j}^z { \hat \tau}_{l}^z\right)  - h\sum_j { \hat \tau}_j^z.
    \label{eq:tpham0}
\end{eqnarray}
The $S=1/2$ pseudospin operators ${\hat{\bm{\tau}}}_j$ here act on the two-dimensional low-energy space of each dimer, where the eigenstates of ${ \hat \tau}^z_j$ with eigenvalues $1/2$ and $-1/2$ correspond, respectively, to the $S^z = 1$ triplet and the singlet states of the full dimer Hilbert space. 
$\langle j,l \rangle_1$ ($ \langle j,l \rangle_2$) denotes pairs of nearest (next nearest) neighbors respectively. The effective exchange anisotropies and pseudo-magnetic field of the low-energy model are:
 \begin{eqnarray}
    \Delta_\nu   = \frac{J^{+}_\nu}{2J^{-}_\nu}, \quad h &=& g \mu_B B \;  - \; J \; -\; 3 (J_1^{+} + J^{+}_2).
    \label{eq:Deltanu}
\end{eqnarray}
The dimensionless parameter $\alpha$ defined by
\begin{eqnarray}
    J^{-}_1 &=& - \frac{\alpha J}{2}, \quad \quad J^{-}_2 = \frac{ \alpha J}{1+\sqrt{5}},
    \label{eq:alpha_def}
\end{eqnarray}
controls the validity  of the perturbation theory, while the ratio $J^{-}_2/|J^{-}_1| = 2/(1+\sqrt{5})$  is tuned to make the ordering wave vector commensurate with a magnetic unit cell of linear dimension $L=5$. 

The commensurate choice $L=5$ is made for computational convenience
and fixes the skyrmion-crystal period in the finite-size calculation.
The stabilization mechanism does not rely on this precise
commensurability, but on the presence of six symmetry-related
finite-wave-vector minima together with the effective easy-axis
anisotropy. We therefore expect small changes in
$J_2^-/|J_1^-|$ to modify the ordering wave vector and shift the phase
boundaries without qualitatively eliminating the SkX phases.
We consider the case $\Delta_1 = \Delta_2 = \Delta$ and find that skyrmions are stabilized only for $\Delta > 1$, consistent with the known requirement of easy-axis anisotropy for skyrmion stabilization in centrosymmetric materials~\cite{Leonov2015}. Notably, the expression for $\Delta_\nu$ given in Eq.~\eqref{eq:Deltanu} reveals an interesting feature: the effective exchange anisotropy is governed by the competition (frustration) between the {\em isotropic} exchange interactions $J_{p \nu}^\prime$ and $J_{c \nu}^\prime$.

In particular, $\Delta > 1$ stabilizes two magnetic-field-induced skyrmion crystal phases, SkX-I and SkX-II, consistent with the known phase diagram of the classical limit of $\tilde{\mathscr{H}}$~\cite{Leonov2015} for values of $\alpha\ll1$. Notably, the skyrmion crystal phase SkX-II, which emerges just below the saturation field, spans a wide range of $\alpha$ values, extending well beyond the perturbative regime of weak inter-dimer coupling. To reduce the number of free parameters, we fix $\Delta = 1.2$ and compute the phase diagram as a function of $B$ and $\alpha$. We employ a magnetic unit cell of linear dimension $L=5$, constructing a finite lattice of up to $2L \times 2L = 4L^2$ dimers with periodic boundary conditions.

\begin{figure*}
    \centering
    \includegraphics[width=1\linewidth]{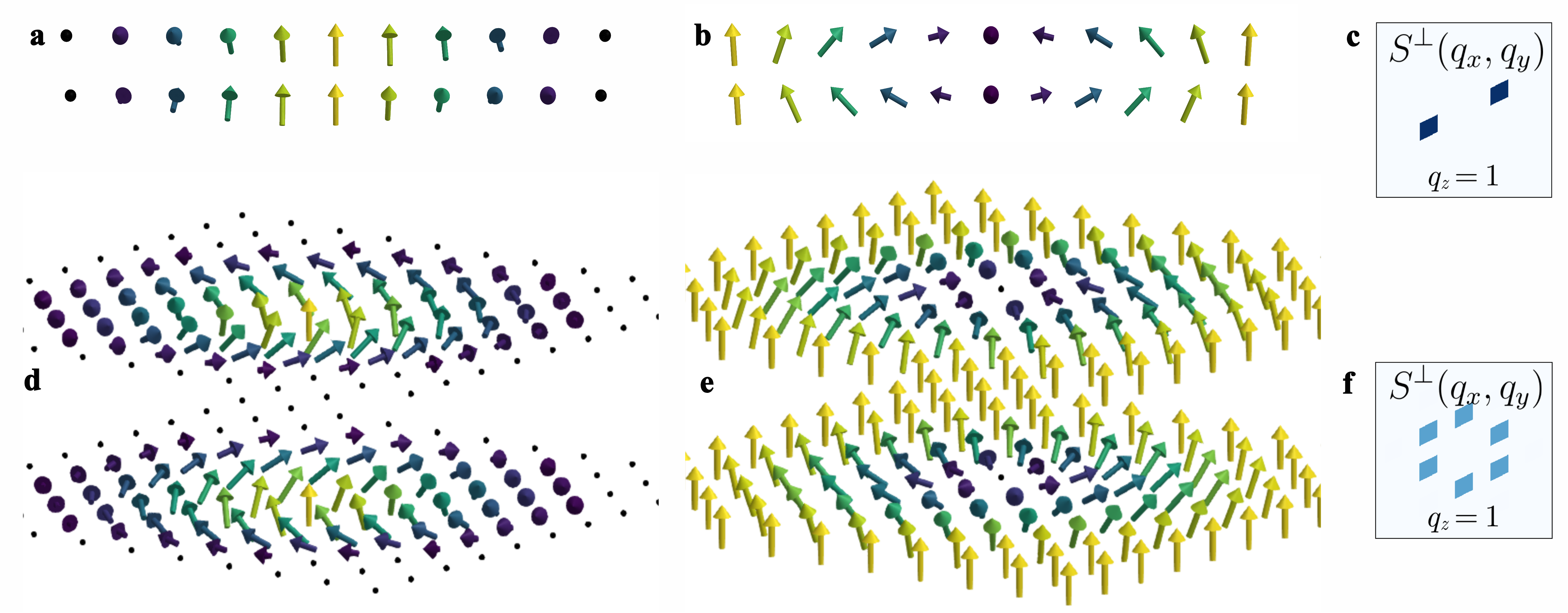}
    \captionsetup{justification=Justified}
    \caption{{\bf Triple-$\bm{Q}$ skyrmions and their single-$\bm{Q}$ cross sections.} {\bf a, b}. Real space distribution of the dipolar sector and {\bf c}. static structure factor for $\mathbb{CP}^3$ single-$\bm{Q}$ spirals. {\bf d, e}.  Real space distribution of the dipolar sector and {\bf f}. triple-$\bm{Q}$ static structure factor for individual $\mathbb{CP}^3$ skyrmions from the phases SkX-I and SkX-II, respectively.}
    \label{fig:skyrm_xsec}
\end{figure*}

The phase diagram includes two zero-field phases: the quantum paramagnet (QPM) and the antiferromagnetic conical spiral (AFM-CS). In the QPM phase, every dimer is in a singlet state. See the next section for a discussion on the spin waves produced in this phase and what can be learned from them. The AFM-CS phase is single-{$\bm Q$} and includes a spiral ordering in each layer such that the $x$ and $y$ components of the dipole moments are antiferromagnetically aligned. At zero field (see blue line in Fig.~\ref{fig:phasediagram}), there is SU(2) symmetry and the $z$ components are also AFM-aligned (see inset); this means that the polarization plane can have any orientation. For $\alpha\gtrsim0.3,$ the ground state configuration has approximately uniform distribution of dipole magnitudes, but the magnitudes are nearly zero. As $\alpha$ increases, the dipole moments retain uniform distribution but increase in magnitude. With applied field, only U(1) symmetry remains, and the spirals are polarized in the $ab$ plane, with ferromagnetic alignment of the $z$ components. As the applied field increases, the dipoles tilt more towards the $\bm{c}$-axis. In addition to the AFM-CS phase, the phase diagram features two more single-{$\bm Q$} phases: the YZ-spiral (YZ-S) and the ferromagnetic conical spiral (FM-CS) (see Appendix~\ref{app:phasechar} for a detailed description). Consider first the YZ-spiral, which serves as the $\mathbb{CP}^3$ analog of a proper screw spiral. As illustrated in Fig.~\ref{fig:skyrm_xsec}a, a single YZ-spiral involves a transverse rotation of the dipole moments, accompanied by a periodic modulation of their magnitudes. The spiral starts from a singlet state, rotates in the $yz$ plane while the dipole magnitude increases until the dimer is in an $S^z=1$ triplet at the midpoint, and then continues rotating and shrinking back down to a singlet state. Notably, the dipoles in the bottom layer rotate in the opposite direction to those in the top layer. In contrast, the AFM-CS (at finite field) and FM-CS phases are characterized by conical spirals in which the $z$ components of the dipoles are FM-aligned, while the $x$/$y$ components are either FM-aligned (FM-CS) or AFM-aligned (AFM-CS). In both conical spiral phases, the dipole magnitude remains constant throughout.

The $\mathbb{CP}^{3}$ skyrmion crystal phases correspond to triple-${\bm Q}$ orderings, meaning they are formed by the superposition of three generalized (single-${\bm Q}$) spirals. These spirals are generated by applying a combined operation: a spatial translation by ${\bm r}$ followed by an SU(4) rotation with angle ${\bm Q} \cdot {\bm r}$ to a uniform reference state (a twist). This transformation differs from a conventional rotation. To see this, consider the regime of weak inter-dimer coupling where the generator of
rotation takes the form ${\hat n} \cdot {\hat{{\bm \tau}}}_j$. Here, the unit vector ${\hat n}$ defines the polarization plane of the generalized spiral, and ${\hat{{\bm \tau}}}_j$ is a vector of the pseudospin operators of the low-energy model Eq.~\eqref{eq:tpham0}. As shown in Appendix~\ref{app:LowEnergyLattice}, when this generator is reexpressed in the Hilbert space of the original Hamiltonian 
it becomes a linear combination of the local SU(4) generators ${ \hat T}^{\mu}_j$ introduced in Eq.~\eqref{eq:basis}, including nonzero contributions from components with $\mu > 6$. It therefore generates a nontrivial SU(4) rotation that acts on both spins of each dimer.

As illustrated in panels {\bf a} and {\bf b} of Fig.~\ref{fig:skyrm_xsec}, this structure leads to a simultaneous rotation and amplitude modulation of the dipole moments, reflecting the alternating singlet and triplet character of the state. A superposition of three such spirals, oriented along directions separated by $120^\circ$, gives rise to the two distinct skyrmion crystal phases: the SkX-I phase, characterized by a triangular lattice of skyrmions with fully polarized cores embedded in a singlet background, and the SkX-II phase, in which the ordering is inverted---singlet cores are surrounded by a fully polarized background (see Fig.~\ref{fig:SkX}).
Remarkably, the latter phase, which emerges just below the fully saturated state, remains stable even when the inter-dimer exchange becomes comparable to the intra-dimer exchange ($\alpha \lesssim 0.5$).

To understand how these novel skyrmions relate to the more familiar variety, we note that one pole of a skyrmion here corresponds to the $S^z = 1$ triplet state, while the other corresponds to the singlet state. For small values of $\alpha$, all physical states of the dimers on the triangular lattice are linear combinations of these two states, forming a non-standard Bloch sphere embedded in $\mathbb{CP}^3$. These kets are in one-to-one correspondence with the three expectation values of the generators ${\hat{{\bm \tau}}}_j$ of an $\mathfrak{su}(2)$ subalgebra of $\mathfrak{su}(4)$. The resulting $\mathbb{CP}^3$ skyrmions are therefore still mappings from $\mathbb{R}^2 \cup \{\infty\} \simeq S^2$ to the sphere $S^2$, as in the familiar case; here, however, the target sphere is the generalized Bloch sphere spanned by the singlet and $S^z = 1$ triplet states rather than the familiar Bloch sphere of purely dipolar states (see Appendix ~\ref{app:SUN}). Analogous to how a standard $\mathbb{CP}^1$ skyrmion has a topological charge given by the number of times the magnetization field wraps $\mathbb{R}^2 \cup \{\infty\}$, the topological charge of the $\mathbb{CP}^3$ skyrmion is determined by how many times the ${\hat{{\bm \tau}}}$-field wraps $\mathbb{R}^2 \cup \{\infty\}$ (see Appendix ~\ref{app:SkDensity}). 
To quantify the extent to which the numerically obtained skyrmions
explore the full $\mathbb{CP}^{3}$ manifold, we calculate
\begin{equation}
w_{\perp,j}
=
1
-
|\langle S=0|Z_j\rangle|^2
-
|\langle S=1,S^z=1|Z_j\rangle|^2
\nonumber
\end{equation}
on every dimer. This quantity measures the local weight outside the
two-dimensional subspace spanned by the singlet and the $S^z=1$
triplet. We focus on SkX-II because this phase extends to the largest
values of $\alpha$ and therefore provides the most stringent test of
the embedded-$\mathbb{CP}^{1}$ description. Table~\ref{tab:coherent_character}
reports the maximum value of $w_{\perp,j}$ over the magnetic unit
cell at several representative points within this phase. The maximum
increases from $0.0012$ at $\alpha=0.10$ to $0.0387$ at
$\alpha=0.44$. Thus, even at the largest value of $\alpha$ reported
in the table, more than $96\%$ of the local-state weight lies within
the singlet--$S^z=1$ triplet subspace on every dimer.

These results show that, although the variational calculation is
performed in the full $\mathbb{CP}^{3}$ manifold, the optimized
SkX-II textures remain quantitatively close to an embedded
$\mathbb{CP}^{1}$ texture throughout the investigated stability
region.

\begin{table}[htbp] 
\caption{
Maximum local weight outside the embedded $\mathbb{CP}^{1}$
subspace at representative points within the SkX-II phase.
The results were obtained using a magnetic unit cell of linear
dimension $L=5$, with $\Delta=1.2$ and
$J_2^-/|J_1^-|=2/(1+\sqrt{5})$.
}
\label{tab:coherent_character} 
\begin{tabular}{ccc} \hline $\alpha$ & $B/J$ & $\max_j\{w_{\perp,j}\}$ \\ \hline 0.10 & 0.89 & 0.0012 \\ 0.16 & 0.82 & 0.0035\\ 0.22 & 0.76 & 0.0065\\ 0.30 & 0.69 & 0.0140\\ 0.36 & 0.65 & 0.0216\\ 0.40 & 0.62 & 0.0289\\ 0.44 & 0.59 & 0.0387\\ \hline \end{tabular}
\end{table}

Using the lattice definition in Eq.~\eqref{eq:totsky}, with the
oriented-plaquette convention $j\to k\to l$ adopted there, we find
that both SkX-I and SkX-II have
\[
|C|=1
\]
per magnetic unit cell. For a magnetic field along the
$\hat{\bm{z}}$ direction, the Hamiltonian is invariant under the
antiunitary transformation
\[
\mathcal{A}=R_{\hat{\bm{e}}_\perp}(\pi)\mathcal{T},
\]
where $\mathcal{T}$ denotes time reversal and
$R_{\hat{\bm{e}}_\perp}(\pi)$ is a global spin rotation by $\pi$
about any axis $\hat{\bm{e}}_\perp$ perpendicular to the field.
For example, choosing $\hat{\bm{e}}_\perp=\hat{\bm{x}}$, this
transformation acts on the spin components as
\[
(S^x,S^y,S^z)\longrightarrow(-S^x,S^y,S^z).
\]
It preserves both the isotropic exchange interactions and the
Zeeman coupling, while reversing the real-space Berry curvature
and hence the sign of the topological charge. Consequently, for a
fixed plaquette orientation, each SkX phase has two degenerate
symmetry-related domains with $C=+1$ and $C=-1$. Once a particular
domain is selected, its integer charge remains unchanged throughout
the corresponding stability region. Thus, although SkX-I and
SkX-II have markedly different local distributions of the
$\mathbb{CP}^{3}$ skyrmion density, they carry the same absolute
topological charge per magnetic unit cell. Independently, reversing
the convention used to orient the triangular plaquettes reverses
the reported sign of $C$.

Besides the single-{$\bm Q$} and skyrmion phases, there are also three double-{$\bm Q$} and two triple-{$\bm Q$} phases. The double-{$\bm Q$} phases exhibit a striped pattern in their scalar chiralities. For $2{\bm Q}$-I and $2{\bm Q}$-II there is little modulation in dipole moment, whereas $2{\bm Q}$-III exhibits significant modulation. The two triple-{$\bm Q$} phases exhibit a checkerboard pattern in their scalar chiralities. The $3{\bm Q}$-I phase exhibits little modulation in dipole moment, whereas the $3{\bm Q}$-II phase has strong dipole moment fluctuation and forms vortex structures that have some resemblance to the SkX-II skyrmions. For more detailed treatment of these phases, see Appendix ~\ref{app:phasechar}.

\begin{figure}
\centering
\includegraphics[width=1.0\linewidth]{Fig4_new.pdf}
\captionsetup{justification=Justified}
\caption{{\bf Real space distribution of the dipolar sector of the $\mathbb{CP}^3$ skyrmion crystals SkX-I and SkX-II.} The length of the arrow represents the magnitude of the dipole moment,
$|\langle \hat{\bm{\mathrm{S}}}_j ^{\sigma}\rangle|=\sqrt{(n_j^1)^2+(n_j^2)^2+(n_j^3)^2}$.  The dipolar textures of the top and bottom layers are shown separately
and explicitly labeled to make their relative orientations clear. The color scale of the arrows indicates $\langle \hat{S}^{z \sigma}_j  \rangle= n_j^3$. The insets display the static spin structure factors $\mathcal{S}^{\perp}(\bm{\mathrm{q}})= \langle n^1_{\bm{\mathrm{q}}}n^1_{-\bm{\mathrm{q}}}+n^2_{\bm{\mathrm{q}}}n^2_{-\bm{\mathrm{q}}} \rangle$ and $\mathcal{S}^{zz}(\bm{\mathrm{q}})= \langle n^3_{\bm{\mathrm{q}}}n^3_{-\bm{\mathrm{q}}} \rangle$, with $\bm{\mathrm{n}}_{\bm{\mathrm{q}}}= \sum_{j} e^{\iu \bm{\mathrm{q}} \cdot \bm{\mathrm{r}}_j} \bm{\mathrm{n}}_j/L$.  The $\mathbb{CP}^3$ skyrmion density distribution $\rho_{j k l}$ [see Eq.~\eqref{eq:totsky}] is indicated by the color of the triangular plaquettes. 
The integrated $\mathbb{CP}^{3}$ topological charge per magnetic
unit cell is $C=+1$ for SkX-I and $C=-1$  for SkX-II, according to the
oriented-plaquette convention of Eq.~\eqref{eq:totsky}.
}
\label{fig:SkX}
\end{figure}

\subsection{Generalized Spin Wave Theory}

The emergence of  field-induced $\mathbb{CP}^3$ SkX phases in a coupled-dimer model with isotropic exchange interactions raises an important question: what simple principles should guide the search for materials capable of hosting such exotic phases? 

According to the phase diagram shown in Fig.~\ref{fig:phasediagram}, the ground state in absence of magnetic field can either correspond to a quantum paramagnetic state, which preserves the SU(2) invariance of the model,  or to the AFM-CS phase consisting of a regular spiral ordering in each layer shown in the inset. Note that the two spirals are anti-aligned due to the large AFM intra-dimer exchange $J$ and that the polarization plane of the spiral is arbitrary because of the SU(2) invariance of ${\hat{\mathscr{H}}}(B=0)$. If a neutron diffraction measurement reveals the latter case, this generalized spiral ordering should be interpreted as a first indicator of a potential field-induced $\mathbb{CP}^3$ SkX phase, which can be regarded as a superposition of three spirals with propagation vectors ${\bm Q}_1$, ${\bm Q}_2$ and ${\bm Q}_3$ (triple-${\bm Q}$ ordering). However, a neutron diffraction experiment does not help if the material is in the QPM state. In this case, only an inelastic neutron scattering experiment can reveal the potential for field-induced skyrmion crystals.

Specifically, the collective modes of the QPM phase, where every dimer is in a singlet state, are triplons with dispersion relation: 
\begin{eqnarray}
    \omega(\bm{k}) &=&\sqrt{J [J +  J^{-}_1\gamma_1(\bm{k}) + J^-_2\gamma_2(\bm{k})]}  \nonumber  \\
    \gamma_1(\bm{k}) &=& \cos{k_x} + 2\cos{\frac{k_x}{2}}\cos{\frac{\sqrt{3}k_y}{2}} \nonumber \\
    \gamma_2(\bm{k}) &=& \cos{\sqrt{3}k_y} + 2\cos{\frac{3k_x}{2}}\cos{\frac{\sqrt{3}k_y}{2}}.
    \label{PM_dispersion}
\end{eqnarray}
The proximity to the AFM-CS phase, driven by increasing the inter-dimer interaction $\alpha$, is signaled by the emergence of six degenerate minima in the singlet-triplon dispersion, as shown in the inset of Fig.~\ref{fig:phasediagram}.  
The continuous transition into the AFM-CS phase occurs when these six modes soften and become gapless with increasing $\alpha$.
These six minima around the local maximum at the $\Gamma$-point are also the indicator of proximity to field-induced SkX phases.

However, our calculations reveal that a second condition must be
satisfied for the stabilization of field-induced SkX phases: the
dimensionless parameter
$\Delta=\Delta_1=\Delta_2$, defined by
\begin{equation}
\Delta_\nu
=
\frac{J_\nu^+}{2J_\nu^-}
=
\frac{J'_{p,\nu}+J'_{c,\nu}}
{2\left(J'_{p,\nu}-J'_{c,\nu}\right)}, 
\end{equation}
must exceed one. The origin of this requirement becomes more
transparent in the limit of weak inter-dimer coupling,
$\alpha\ll 1$, where $\Delta>1$ indicates that the effective
pseudospin-$1/2$ model~\eqref{eq:tpham0} exhibits easy-axis exchange
anisotropy. This anisotropy is a well-established prerequisite for
stabilizing field-induced SkX phases in centrosymmetric
magnets~\cite{Leonov2015}.

While the value of $\Delta$ cannot be extracted from fitting the triplon dispersion which depends only on the differences $J^{-}_{\nu}$, it can be extracted from fitting the dipersion of the magnon modes in the FP phase. Since the Hamiltonian is invariant under a global symmetry operation that exchanges the two spins of each dimer, the parity of the magnon modes under this operation is good quantum number. The dispersion relations of the symmetric $(+)$ and anti-symmetric $(-)$ modes are $\omega_+(\bm{k})= \epsilon_+(\bm{k})$ and 
$\omega_-(\bm{k})= \epsilon_-(\bm{k})-J$, with
\begin{eqnarray}
    \epsilon_{\pm}(\bm{k})&=&g\mu_BB + J^{\pm}_{1} \gamma_1(\bm{k}) +  J^{\pm}_{2} \gamma_2(\bm{k}) - 6 (J^{\pm}_{1} + J^{\pm}_{2}) .
    \label{magnon+}
\end{eqnarray}

\begin{figure*}
    \centering
    \includegraphics[width=1\linewidth]{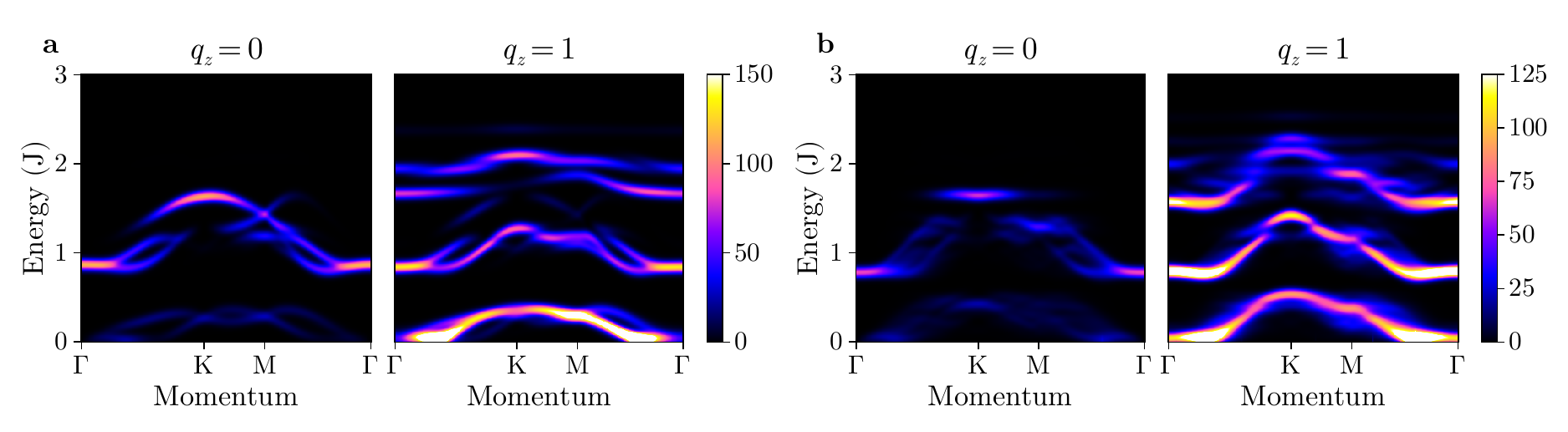}
    \captionsetup{justification=Justified}  
    \caption{{\bf Magnetic excitations of the YZ-Spiral and SkX-I Phases.} {\bf a}. The dynamical spin structure factor (DSSF) of a system in the YZ-Spiral phase calculated with linear spin wave theory (LSWT) at the point $(\alpha,B)=(0.1,0.865)$ of the phase diagram of Fig.~\ref{fig:phasediagram}. {\bf b}. The DSSF of a system in the SkX-I phase calculated using LSWT at the point $(\alpha,B)=(0.14,0.779)$ of the phase diagram in Fig.~\ref{fig:phasediagram}. Each pair of panels shows the $q_z=0$ (left) and $q_z=1$ channels (right). Calculations were performed using the \texttt{Sunny.jl} package and include artificial Gaussian broadening with $\sigma=0.042$ J.}
    \label{fig:SWT}
\end{figure*}
The final step, after verifying the conditions for the emergence of $\mathbb{CP}^3$ skyrmions under a magnetic field, is to experimentally identify
the SkX phases using a combination of diffraction and inelastic neutron scattering (NS) measurements. As in the case of conventional
skyrmion phases, a major challenge for neutron diffraction studies is to distinguish an intrinsic triple-${\bm Q}$ ordering -- characterized by
six diffraction peaks at $\pm {\bm Q}_{\nu}$, as shown in the insets of Fig.~\ref{fig:skyrm_xsec} -- from a multi-domain structure of single ${\bm Q}$ orderings, in which each domain contributes to the intensity of a specific pair of peaks $\pm {\bm Q}_{\nu}$.
As recently demonstrated in Ref.~\cite{Park25}, differences in the inelastic NS spectra between triple-${\bm Q}$ and single-${\bm Q}$ phases can be used to discriminate between the two states. Motivated by this work, we computed the DSSF for the competing YZ-Spiral and SkX-I phases using the software package \texttt{Sunny.jl}~\cite{Dahlbom2025}. As shown in Fig.~\ref{fig:SWT}, there are some clear differences seen between the YZ-Spiral and SkX-I phases. In particular, in the anti-symmetric ($q_z=1$) channel, three bands of the SkX-I phase have similar intensities, while the lowest energy band has much greater intensity for the YZ-spiral phase.
Also, following Ref.~\cite{Park25}, triple-$\bm{Q}$ and multi-domain single-$\bm{Q}$ can be distinguished through the dispersion cones (associated with the Goldstone modes) at the ordering wave-vectors  which in the former case have circular cross-sections and in the latter have elliptical cross-sections. See Fig.~\ref{fig:constant_energy_slices} in Appendix \ref{app:DSSFextras}.

We finally observe that the SkX-I and SkX-II phases exhibit a net
scalar spin chirality on both layers,
\[
\langle \boldsymbol{\hat{S}}_{j+} \rangle
\cdot
\left(
\langle \boldsymbol{\hat{S}}_{k+} \rangle
\times
\langle \boldsymbol{\hat{S}}_{l+} \rangle
\right)
=
\langle \boldsymbol{\hat{S}}_{j-} \rangle
\cdot
\left(
\langle \boldsymbol{\hat{S}}_{k-} \rangle
\times
\langle \boldsymbol{\hat{S}}_{l-} \rangle
\right),
\]
because the $z$ components of the two spins in each dimer are
parallel, whereas their $x$ and $y$ components are antiparallel.
This ordinary scalar spin chirality is an additional dipolar
signature of the noncoplanar SkX textures, but it is not, in
general, equal to their $\mathbb{CP}^{3}$ topological density.
In the weak-coupling regime, where the dimer states remain within
the $\mathbb{CP}^{1}$ subspace spanned by the singlet and the
$S^z=1$ triplet, the integrated layer spin chirality is related
to the winding number of the corresponding pseudospin texture.
Once the texture explores the full $\mathbb{CP}^{3}$ manifold,
however, the topological density also contains contributions from
the remaining bilinear and multipolar components of the color
field. The ordinary scalar spin chirality is then neither
quantized nor sufficient to determine the topological charge $C$.

More generally, electronic or phononic degrees of freedom coupled
to the spatially varying dimer texture can acquire momentum-space
Berry curvature in the presence of a periodic
$\mathbb{CP}^{3}$ skyrmion crystal. The magnitude, sign, and
momentum dependence of the induced Berry curvature depend on the
microscopic coupling between these degrees of freedom and the
underlying spin texture, as well as on the resulting electronic
or phononic band structure. The corresponding electronic Hall,
magneto-optical, or phonon thermal Hall responses are therefore
material-dependent and need not be quantized. In particular, the
nonzero scalar spin chirality may provide an additional indirect
signature through magnetic circular dichroism
(MCD)~\cite{Feng2020}, but it should not be interpreted as a
direct measurement of the $\mathbb{CP}^{3}$ topological charge.

\subsection{Single-skyrmion solutions}

In addition to the two skyrmion-crystal phases SkX-I and SkX-II, we
find metastable isolated skyrmions with the corresponding core and
background structures. Starting from spatially localized initial
configurations, we minimize the full $\mathrm{SU}(4)$ coherent-state
energy using the conjugate-gradient procedure described in
Sec.~V. The resulting configurations converge to spatially localized
stationary solutions and remain unchanged under further energy
minimization, demonstrating their local stability within our
numerical procedure.

\begin{figure*}
\centering
\includegraphics[width=\linewidth]{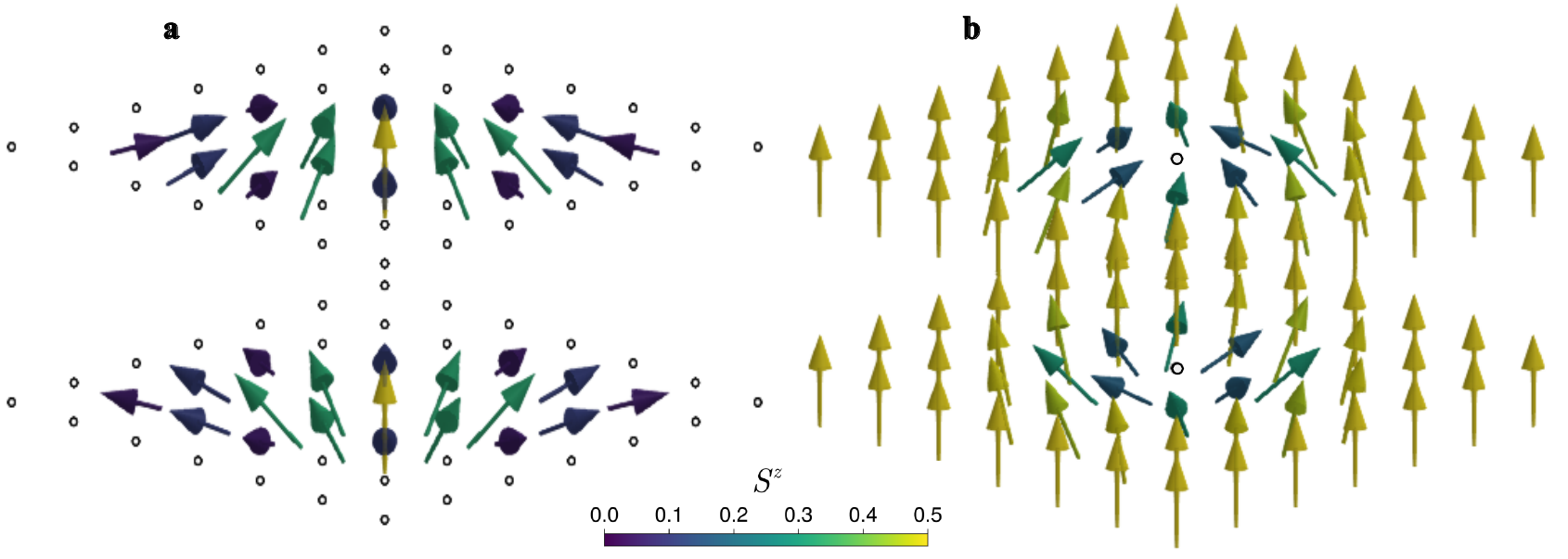}
\captionsetup{justification=raggedright,singlelinecheck=false}
\caption{
\textbf{Single-skyrmion solutions.}
\textbf{a.} Metastable SkX-I-type skyrmion inside the QPM phase
at $(\alpha,B/J)=(0.14,0.74)$, with an $S^z=1$ triplet core
embedded in a singlet background.
\textbf{b.} Metastable SkX-II-type skyrmion inside the fully
polarized phase at $(\alpha,B/J)=(0.24,0.79)$, with a singlet
core embedded in a fully polarized background.
Both calculations shown here were performed on $L\times L$ dimer
lattices with periodic boundary conditions using $L=10$.
Finite-size effects are negligible for $L\gtrsim 7$.
The model parameters are $J_1^-=-\alpha J/2$,
$J_2^-=\alpha J/(1+\sqrt{5})$, and $\Delta=1.2$.
Arrow length represents the dipole magnitude, arrow color represents
its $z$ component, and open circles represent singlet dimers.
}
\label{fig:metastable}
\end{figure*}

Figure~\ref{fig:metastable}(a) shows an SkX-I-type skyrmion inside
the QPM phase at $(\alpha,B/J)=(0.14,0.74)$. It consists of an
$S^z=1$ triplet core embedded in a singlet background.
Figure~\ref{fig:metastable}(b) shows an SkX-II-type skyrmion inside
the fully polarized phase, above the saturation field, at
$(\alpha,B/J)=(0.24,0.79)$. In this case, the structure is inverted:
a singlet core is embedded in a fully polarized background.

\section{ Discussion \label{sec:Discussion}}

Much of our intuition in condensed matter physics is grounded in semi-classical approaches to quantum mechanical models. These treatments reveal some of the most striking examples of emergence in many-body systems, including the formation of mesoscale particles known as \emph{topological solitons}, which can organize into distinct states of matter---such as crystals, liquids, and gases. These classical states of matter often serve as precursors to more exotic quantum phases, such as \emph{quantum liquids of topological solitons}, whose elementary excitations carry fractional quantum numbers.

The nature of the emergent topological solitons is determined by the phase space of the classical theory, which becomes the target space of the field theory obtained in the long-wavelength limit. As was observed by Perelomov~\cite{Perelomov1972}, quantum mechanical $N$-level systems admit different classical limits with phases spaces corresponding to coherent state manifolds of different Lie algebras. 
The remarkable observation, which applies to spin systems, forces us to reconsider the standard classical limit, based on
the $\mathbb{CP}^1$ manifold of SU(2) coherent states that is typically applied to spin systems~\cite{Hao21,David2022,Dahlbom2022b}.

In a recent study, we demonstrated that the $\mathbb{CP}^3$ manifold of SU(4) coherent states provides the most natural description for the semi-classical dynamics of low-dimensional coupled $S = 1/2$ dimer systems~\cite{David2024}. In this manuscript, we show how frustrated inter-dimer interactions naturally give rise to the emergence of generalized $\mathbb{CP}^3$ skyrmion crystals in equilibrium, as well as metastable single skyrmion solutions. 

The emergence of metastable topological textures within a quantum paramagnetic (QPM) phase, composed of singlets on each dimer, highlights the need to replace the conventional order parameter manifold, absent in the isotropic QPM, with the full phase space manifold. A natural consequence of this shift is that skyrmions, along with other topological textures such as chiral solitons, can span both isotropic regions (with no observable local order parameter) and regions where the local order parameter acquires a finite expectation value. The resulting amplitude modulation of the local order parameter enhances its coupling to spatial gradients of external fields. In particular, the interaction between these generalized skyrmion crystals and the phonon field may give rise to a sizable momentum-space Berry curvature in the phonon bands, potentially leading to the emergence of a phonon thermal Hall effect of topological origin. 

Stacked triangular-lattice materials with strong interlayer interactions---such as Ba$_3$Mn$_2$O$_8$~\cite{stone2008,samulon2009,samulon2010}, $\mathrm{Ba_3CoNb_2O_9}$~\cite{Lee2014,Yokota2014}, and $\mathrm{CsNiCl_3}$~\cite{Johnson1979}---may provide promising platforms for realizing the $\mathbb{CP}^3$ skyrmions proposed in this work. However, unlike the scenario considered here, these materials exhibit magnetic ordering wave vectors near the boundary of the Brillouin zone, suggesting that the corresponding skyrmions would be antiferromagnetic in nature.

\section{Data availability}
All data presented in this study can be reproduced using the code package described in \textbf{Code availability}.

\section{Code availability}
The algorithms used in our numerical simulations are described in the \textbf{Methods} section. The numerical code is implemented in Julia and can be found at \url{https://github.com/fwilli18/2025-Williams-Skyrmions_frustrated_quantum_dimers}.

\section{Acknowledgments}
The work by F.W. was primarily supported by the National Science Foundation Materials Research Science and Engineering Center program through the UT Knoxville Center for Advanced Materials and Manufacturing (DMR-2309083). D.D. S.A. and C.D.B.~acknowledge support from U.S. Department of Energy, Office of Science, Office of Basic Energy Sciences, under Award No.~DE-SC0022311. K.B.  acknowledges the support of the U.S. Department of Energy through the LANL/LDRD Program. 
\footnote{This manuscript has been authored by UT-Battelle, LLC, under contract DE-AC05-00OR22725 with the
US Department of Energy (DOE). The US government retains and the publisher, by accepting the article for
publication, acknowledges that the US government retains a nonexclusive, paid-up, irrevocable, worldwide
license to publish or reproduce the published form of this manuscript, or allow others to do so, for US government
purposes. DOE will provide public access to these results of federally sponsored research in accordance with
the DOE Public Access Plan (https://www.energy.gov/doe-public-access-plan)}

\section{Author contributions}
C.D.B conceived the project. F.W. and D.D. designed the numerical code. F.W. performed the numerical simulations. F.W. and C.D.B. analyzed the simulation results. All authors contributed to the writing of the manuscript.

\section{Competing interests}
The authors declare no competing interests.

\appendix

\section{Spin expectation values for  low-energy SU(2) coherent states}
\label{app:expecvals}

Coherent states of the low-energy classical Hamiltonian \eqref{eq:tpham0} correspond to arbitrary linear combinations of the two states, singlet $\vert S=0\rangle$ and $S^z=1$ triplet $\vert S=1,S^z=1\rangle$ that span the low-energy subspace,
\begin{equation}
\vert\psi\rangle = \cos\frac{\theta}{2}e^{i\phi/2}\vert S=1,S^z=1\rangle + \sin\frac{\theta}{2}e^{-i\phi/2}\vert S=0\rangle.
\end{equation}
It is instructive to examine the expectation values of the two spins that comprise a dimer.
By rewriting these states in terms of the original product basis, it is a straightforward calculation to find $\langle \psi | { \hat S}^\mu_\sigma | \psi \rangle$ for $\mu=x, y, z$ and $\sigma = \pm$:

\begin{equation}
\langle { \hat {\bm S}}_{\pm}\rangle = \left( \frac{\mp 1}{2\sqrt{2}} \sin \theta \cos \phi,\frac{\pm 1}{2\sqrt{2}}\sin \theta \sin \phi, \frac{1}{4} \left( 1 + \cos \theta \right) \right).
\end{equation}
Note that the expectation value of the $xy$ spin components are antiferromagnetically aligned for finite hybridization between the 
singlet and the triplet, while the $z$-components are parallel for $\theta\neq \pi$.

We also compute the expectation values of the remaining nine SU(4) observables with respect to the low-energy coherent states:
\begin{equation}
\begin{split}
    \left\langle {\hat {\bm S}}_{+} \cdot { \hat {\bm S}}_{-}  \right\rangle \quad &= \quad \frac{1}{4} \left( \cos^2 \frac{\theta}{2} - 3 \sin^2 \frac{\theta}{2} \right) \\
    \left\langle { \hat {\bm S}}_{+} \times { \hat {\bm S}}_{-}  \right\rangle \quad &= \quad \frac{1}{2 \sqrt{2}} \left( \sin\theta \sin\phi,\sin\theta \cos\phi, 0 \right)\\
    Q^{\mu \mu} \quad &= \quad \left(\frac{-1+3\delta_{\mu,z}}{6} \right)\cos^2 \frac{\theta}{2} \\
    Q^{\mu \nu} \quad &= \quad 0 \quad \left(\mu \neq \nu \right).\\
\end{split}
\end{equation}

\section{Triangular Lattice pseudospin Model}
\label{app:LowEnergyLattice}
\subsection{Mapping the pseudospin model to the full Hamiltonian}
We now focus on the low-energy effective pseudospin model defined by Eqs.~\eqref{eq:tpham0} and~\eqref{eq:Deltanu}, and explicitly map its parameters back to those of the original Hamiltonian $\hat{\mathscr{H}}$. This is achieved by inverting Eqs.~\eqref{eq:Deltanu} and~\eqref{eq:alpha_def}, while keeping the ratio $\frac{J_2^-}{J_1^-}$ fixed. The value of this ratio is chosen such that the classical magnetic ground state exhibits a magnetic unit cell of length 5 (as will be explained at the end of this appendix). To parameterize the intra- and inter-dimer couplings, we set
\[
J_1^- = -\frac{\alpha J}{2}, \quad J_2^- = \frac{\alpha J}{1 + \sqrt{5}}.
\]
We consider the case of easy-axis anisotropy, where
\[
\Delta_1 = \Delta_2 = \Delta > 1.
\]

Using Eqs.~(3) and (21), the effective exchange anisotropy is
related to the microscopic couplings through
\[
\Delta_\nu
=
\frac{J_\nu^+}{2J_\nu^-}
=
\frac{J'_{p,\nu}+J'_{c,\nu}}
{2\left(J'_{p,\nu}-J'_{c,\nu}\right)}.
\]

The Hamiltonian parameters of the full model, given in terms of the pseudospin model parameters, are then,
\begin{eqnarray}
    J^\prime_{p,\nu} &=& \left(\Delta + 1/2 \right)\alpha J \left( -\delta_{\nu,1} + \frac{2\delta_{\nu,2}}{1+\sqrt{5}} \right), \nonumber \\
    J^\prime_{c,\nu} &=& \left(\Delta - 1/2 \right)\alpha J \left(-\delta_{\nu,1} + \frac{2\delta_{\nu,2}}{1+\sqrt{5}} \right), \nonumber \nonumber \\
    g \mu_B B &=& h + J + 3J \alpha \Delta  \left(\frac{1-\sqrt{5}}{1+\sqrt{5}} \right).
    \label{fullparam}
\end{eqnarray}
We limit consideration to the case where \( J_1^- < 0 \) and \( J_2^- > 0 \). This choice implies that the exchange interactions \( J'_{p1} \) and \( J'_{c1} \) are ferromagnetic (\( J'_{p1}, J'_{c1} < 0 \)), while \( J'_{p2} \) and \( J'_{c2} \) are antiferromagnetic (\( J'_{p2}, J'_{c2} > 0 \)).

To appreciate the nature of spirals in the full SU$(4)$ model, we express the pseudospin operators in terms of the generators of SU$(4)$ in the basis given in Eq.~\eqref{eq:basis}:
\begin{eqnarray}
    {\hat{\tau}}_j^x&=&\frac{1}{2\sqrt{2}} \left( -{ \hat{T}}^1+{ \hat{T}}^4-{\hat{T}}^{13}+{ \hat{T}}^{14} \right) \nonumber \\
    {\hat{\tau}}_j^y&=&\frac{1}{2\sqrt{2}} \left( -{\hat{T}}^2+{\hat{T}}^5-{\hat{T}}^{10}+{\hat{T}}^{11} \right) \\
    {\hat{\tau}}_j^z&=&\frac{1}{4} \left( -2 {\hat{I}}+ { \hat{T}}^3+{\hat{T}}^6+2{ \hat{T}}^7+2{\hat{T}}^8+6{\hat{T}}^9 \right). \nonumber
    \label{eq:pseudo_spins_in_full_basis}
\end{eqnarray}

\subsection{Momentum Space Hamiltonian}
Next, we rewrite the pseudospin Hamiltonian $\tilde{\mathscr{H}}$ in momentum space by Fourier transforming the pseudospin operators:
\begin{eqnarray}
\tilde{\bm \tau}^{\phantom{\dagger}}_{\bm k}   &=& \frac{1}{\sqrt{N}} \sum_{j} e^{i {\bm k} \cdot {\bm r}_j} {\bm \tau}^{\phantom{\dagger}}_j, 
\nonumber \\
{\bm \tau}^{\phantom{\dagger}}_{j}   &=& \frac{1}{\sqrt{N}} \sum_{\bm k} e^{-i {\bm k} \cdot {\bm r}_j} \tilde{\bm \tau}^{\phantom{\dagger}}_{\bm k}.
\label{Eq:FT}
\end{eqnarray}
Note that $\tilde{\bm \tau}^*_{\bm k} = \tilde{\bm \tau}^{\phantom{\dagger}}_{-{\bm k}}$ because ${\bm \tau}^{\phantom{\dagger}}_j$ is real ${\bm \tau}^{\phantom{\dagger}}_j = {\bm \tau}^*_j$. The expression for the pseudospin Hamiltonian $\tilde{\mathscr{H}}$ in momentum space is
\begin{equation}
\tilde{\mathscr{H}}= \sum_{\bm k} {\cal J}({\bm k}) (\tilde{\tau}^x_{\bm k} \tilde{\tau}^x_{-{\bm k}} + \tilde{\tau}^y_{\bm k} \tilde{\tau}^y_{-{\bm k}}+ \Delta \tilde{\tau}^z_{\bm k} \tilde{\tau}^z_{-{\bm k}} ) - \sqrt{N} h \tau^z_{\bm 0} + C,
\label{eq:Heff}
\end{equation}
where
\begin{eqnarray}
        {\cal J}({\bm k}) &=&  J_1^- \left[2\cos{\left(\frac{k_x a}{2}\right)} \cos{\left(\frac{\sqrt{3} k_y a}{2}\right)}+\cos{\left(k_x a\right)}\right] \\
        &+& J_2^- \left[2\cos{\left(\frac{3 k_x a}{2}\right)} \cos{\left(\frac{\sqrt{3} k_y a}{2}\right)}+\cos{\left(\sqrt{3} k_y a\right)}\right]. \nonumber
\end{eqnarray}
By Taylor expanding around ${\bm k}={\bm 0}$, we have
\begin{eqnarray}
    {\cal J}({\bm k}) &=& 
    3(J_1^-+J_2^-) -\frac{3a^2}{4}(J_1^-+3J_2^-) k^2 \nonumber \\
    &+& \frac{3a^4}{64} (J_1^-+9 J_2^-)k^4 \label{JTexp} \\
    &-& \left(\frac{a}{2}\right)^6 \left( \frac{1}{12} (J_1^-+27J_2^-)  + \frac{1}{5!} (J_1^--27J_2^-) \cos{6 \phi}\right)k^6 \nonumber \\
    &+& {\cal O}(k^8),\nonumber
\end{eqnarray}
where ${\bm k} = k (\cos{\phi},\sin{\phi})$. Note that $-(J_1^--27J_2^-) >0$, implying that the six vectors $\pm{\bm Q}_{\nu}$ ($\nu=1,2,3$) that minimize the dispersion relation have values of $\phi=\pm \phi_{\nu}$ with $\phi_{\nu}=-\pi/6 +  2 \nu \pi/3 $. 

From this point on, we set the lattice constant \( a \) as the unit of length (\( a = 1 \)).  
Equation~\eqref{JTexp} shows that the anisotropy of the hexagonal lattice first appears at sixth order in the expansion. This anisotropy selects optimal wave vectors aligned along the directions connecting the \(\Gamma\) point to the \(\pm M\) points when \( J_2^- \gtrsim |J_1^-|/3 = J_2^c \).

From the quadratic and quartic terms of the expansion, we can determine the magnitude
$|{\bm Q}_{\nu}|$ of the six optimal wave vectors that minimize ${\cal J}({\bm k})$ for $J_2^- \gtrsim |J_1^-|/3$:
\begin{eqnarray}
    |{\bm Q}_{\nu}| \simeq  \sqrt{\frac{8 ( J_2^- -J^c_2 )}
    {3 ( J_2^- - J^c_2/3 )}} \simeq 2 \sqrt{\frac{J_2^- -J^c_2}{J^c_2}}.
\nonumber 
\end{eqnarray}

 The function \( \mathcal{J}(\bm{k}) \) exhibits a global minimum at \( \bm{k} = \bm{0} \) for \( J_2^- \leq |J_1^-|/3 \equiv J_2^c \), signaling ferromagnetic ordering. Beyond this Lifshitz transition, for \( J_2^- \gtrsim J_2^c \), the single minimum at \(\mathbf{0}\) splits into six symmetry-related global minima located at wave vectors \( \pm \mathbf{Q}_\nu \), where the \( \mathbf{Q}_\nu \) vectors are related by successive \(120^\circ\) rotations:
\begin{equation}
{\bm Q}_{1} = Q {\bm b}_1, \quad   {\bm Q}_{2} = Q {\bm b}_2, \quad {\bm Q}_{3} = - Q ({\bm b}_1 + {\bm b}_2).
\end{equation}
The magnitude \( Q \) of the ordering wave vectors is determined by minimizing the function
\begin{eqnarray}
{\cal J}(Q {\bm b}_2) =  {\tilde J_1} [2 \cos{(2\pi Q)}+1 ] 
+ {\tilde J_2} \left[2 \cos{(2\pi Q)}+\cos{(4\pi Q)}\right], \nonumber   
\end{eqnarray}
which yields the condition:
\begin{equation}
\cos(2\pi Q) = -\frac{1}{2} \left( \frac{J_1^-}{J_2^-} + 1 \right).
\label{Q_ordering}
\end{equation}

\begin{figure}
     \centering
     \includegraphics[width=1.0\linewidth]{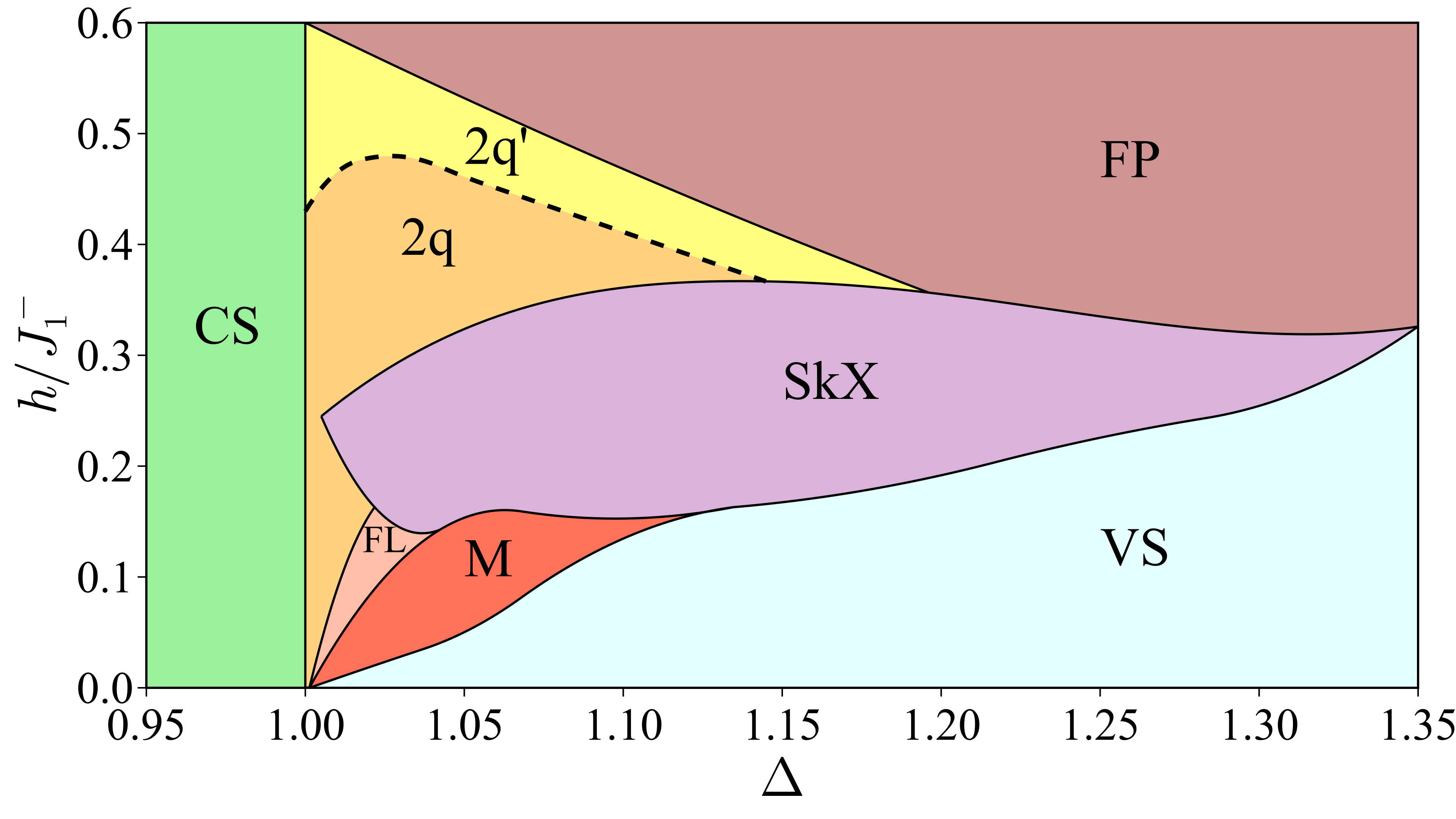}
     \captionsetup{justification=Justified}
     \caption{{\bf Zero temperature phase diagram of low-energy Hamiltonian $\tilde{\mathscr{H}}$.} The diagram is computed in terms of pseudo-magnetic field $h$ and effective anisotropy $\Delta$. FM is ferromagnetic phase; SkX is the skyrmion crystal; VS is the vertical spiral; CS is the conical spiral; FL is the flop state; M is the VS with two in-plane sinusoidal modulations; $2q$ and $2q'$ are the two double-q states. }
     \label{PSPD}
 \end{figure}

Since we are free to choose the relationship between \( J_1^- \) and \( J_2^- \), provided that \( J_2^- > \frac{|J_1^-|}{3} \), we set
\[
J_1^- = -\frac{\alpha J}{2}, \quad J_2^- = \frac{\alpha J}{1 + \sqrt{5}},
\]
which yields
\begin{equation}
    Q = \frac{1}{5},
    \label{eq:QL}
\end{equation}
corresponding to a magnetic unit cell of size \( L = 5 \).
With this choice of parameters, we computed the zero-temperature phase diagram of \( \tilde{\mathscr{H}} \) as a function of the effective field \( h \) and the effective anisotropy \( \Delta \), as shown in Fig.~\ref{PSPD}. This phase diagram qualitatively agrees with the \( T = 0 \) phase diagram obtained for a related XXZ model in which the exchange anisotropy is replaced by a single-ion anisotropy~\cite{Leonov2015}. We selected \( \Delta = 1.2 \) as a representative point lying within an extended skyrmion phase. This choice, together with the mapping defined in Eq.~\eqref{fullparam}, enabled us to compute the full set of microscopic Hamiltonian parameters corresponding to each point in the \( B \) versus \( \alpha \) phase space.

\section{Structure Constants}
\label{app:structure_constants}
In the main text, we expressed the model Hamiltonian in terms of the generators \( \hat{T}^\mu \) of SU(4), defined with respect to a particular basis (Eq.~\eqref{eq:basis}). The structures constants of this basis, $f_{\mu\nu\eta}$, defined by Eq.~\eqref{eq:commutation_relations}, are readily calculated as
\begin{eqnarray}
    f_{1,2,3} &=& f_{1,8,11} = f_{2,9,13} = f_{2,11,14}=f_{3,7,15}=1,\nonumber \\
    f_{1,9,10} &=& f_{1,12,15} = f_{2,7,12} = f_{3,8,14}=f_{3,10,13}=-1,\nonumber \\
    f_{4,5,6} &=& f_{4,8,10} = f_{5,9,12} = f_{5,10,15}=f_{6,7,14}=1,\\
    f_{4,9,11} &=& f_{4,13,14} = f_{5,7,13} = f_{6,8,15}=f_{6,11,12}=-1.\nonumber 
\end{eqnarray}

\section{Hamiltonian Parameters}
\label{app:H_param}
The explicit coefficients of the SU(4) Hamiltonian in the continuum limit, as given in Eq.~\eqref{HSU4cont}, are

\begin{eqnarray}
{\cal B}^\mu &=& \left(\frac{J}{2a^2}+\frac{3}{2}\left(J_{c1}^{\prime} + J_{c2}^{\prime}\right)\right)(\delta_{\mu,7}+\delta_{\mu,8}+\delta_{\mu,9}) \nonumber \\
&-& \frac{g \mu_B B}{a^2}(\delta_{\mu,3}+\delta_{\mu,6}) \nonumber \\
{\cal J}^{\mu \nu}_1 &=& -\frac{3}{8} (J_{p1}^{\prime} + 3 J_{p2}^{\prime})\delta_{\mu,\nu} \left(\sum_{\mu'=1}^{6}\delta_{\mu,\mu'} \right) \nonumber \\
&-& \enspace \; \frac{3}{4} (J_{c1}^{\prime} + 3 J_{c2}^{\prime})\delta_{\mu,\nu-3}\left(\sum_{\mu'=1}^{3}\delta_{\mu,\mu'}\right) \\
{\cal J}^{\mu \nu}_2 &=& \frac{3a^2}{128} (J_{p1}^{\prime} + 9 J_{p2}^{\prime})\delta_{\mu,\nu} \left(\sum_{\mu'=1}^{6}\delta_{\mu,\mu'} \right) \nonumber \\
&+& \frac{3a^2}{64} (J_{c1}^{\prime} + 9 J_{c2}^{\prime}) \delta_{\mu,\nu-3}\left(\sum_{\mu'=1}^{3}\delta_{\mu,\mu'}\right). \nonumber
\end{eqnarray}

\section{$\mathbb{CP}^3$ Phase Characterizations}
\label{app:phasechar}
In the main text, we presented a detailed analysis of the two skyrmion phases (SkX-I and SkX-II) that arise in the classical SU(4) limit of the Hamiltonian model given in Eq.~\eqref{hamil}. Below, we provide characterizations of the remaining phases reported in Fig.~\ref{fig:phasediagram}.

\subsection{Single-Q Spiral Phases}

\begin{figure*}
    \centering
    \includegraphics[width=1\linewidth]{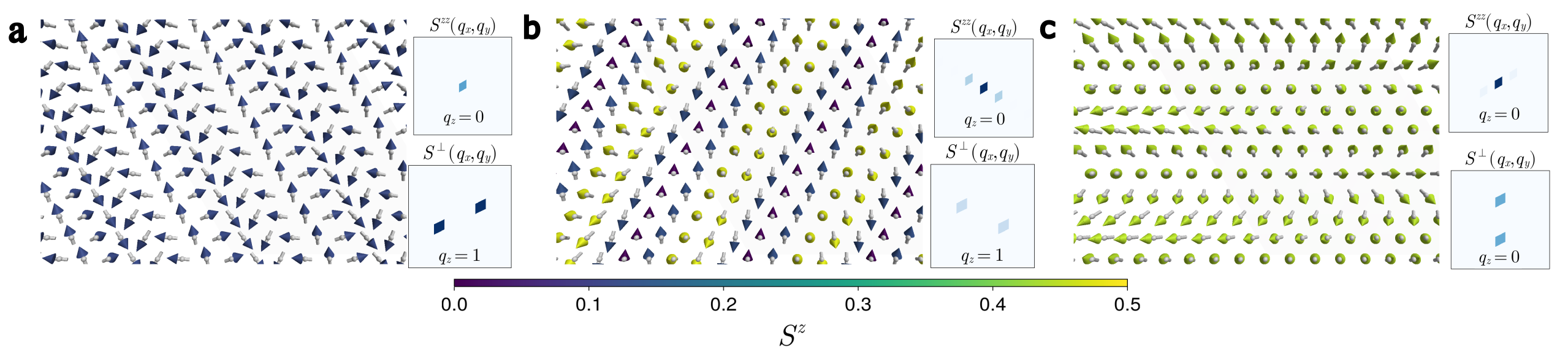}
    \captionsetup{justification=Justified}
    \caption{{\bf Real space distribution of the top layer dipolar sector of the three $\mathbb{CP}^3$ single-{$\bm Q$} orderings.} {\bf a}. AFM-CS. {\bf b}. YZ-S. {\bf c}. FM-CS. The length of the arrow represents the magnitude of the dipole moment of the color field $|\langle \hat{\bm{\mathrm{S}}}_j ^{+}\rangle|=\sqrt{(n_j^1)^2+(n_j^2)^2+(n_j^3)^2}$. The color scale of the arrows indicates $\langle \hat{S}^{z +}_j  \rangle= n_j^3$. The insets display the static spin structure factors $\mathcal{S}^{\perp}(\bm{\mathrm{q}})= \langle n^1_{\bm{\mathrm{q}}}n^1_{-\bm{\mathrm{q}}}+n^2_{\bm{\mathrm{q}}}n^2_{-\bm{\mathrm{q}}} \rangle$ and $\mathcal{S}^{zz}(\bm{\mathrm{q}})= \langle n^3_{\bm{\mathrm{q}}}n^3_{-\bm{\mathrm{q}}} \rangle$, with $\bm{\mathrm{n}}_{\bm{\mathrm{q}}}= \sum_{j} e^{\iu \bm{\mathrm{q}} \cdot \bm{\mathrm{r}}_j} \bm{\mathrm{n}}_j/L$.}
    \label{fig:single-Q}
\end{figure*}

Fig.~\ref{fig:phasediagram} features three distinct single-$ \bm{Q} $ ordered phases. The largest of these is the AFM-CS phase (antiferromagnetic conical spiral; Fig.~\ref{fig:single-Q}a), in which the spins rotate within the \( xy \)-plane. In this phase, the dipole moment exhibits minimal modulation along the spiral.

The second-largest single-\( \bm{Q} \) phase is the FM-CS phase (ferromagnetic conical spiral; Fig.~\ref{fig:single-Q}c). Here, the spins primarily align along the \( z \)-axis and spiral around it, with only a slight modulation of the dipole moment.

The smallest of the single-$\bm{Q}$ phases is the YZ-spiral (Fig.~\ref{fig:single-Q}b), which is analogous to a proper-screw spiral. In this phase, the dipole moments lie predominantly in the $yz$-plane perpendicular to the propagation direction and vary in magnitude. This modulation causes the dimers to continuously evolve from pure singlets to fully polarized $S^z = 1$ triplets and back, in such a way that the dipole moments never acquire components pointing in the negative $z$-direction.

\subsection{Double-Q Phases}
\begin{figure*}
    \centering
    \includegraphics[width=1\linewidth]{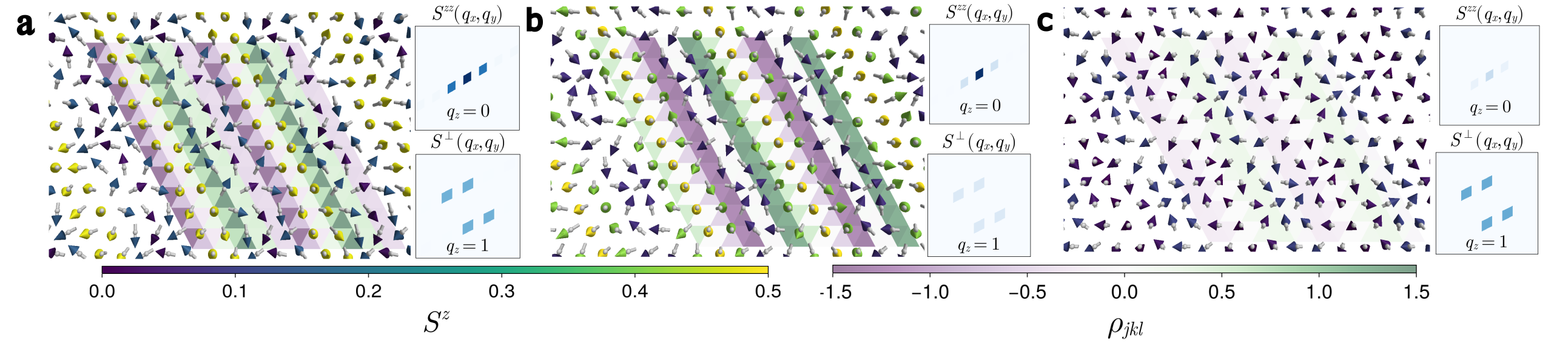}
    \captionsetup{justification=Justified}
    \caption{{\bf Real space distribution of the top layer dipolar sector of the three $\mathbb{CP}^3$ double-Q orderings.} {\bf a}. 2Q-I. {\bf b}. 2Q-II. {\bf c}. 2Q-III. The length of the arrow represents the magnitude of the dipole moment of the color field $|\langle \hat{\bm{\mathrm{S}}}_j ^{+}\rangle|=\sqrt{(n_j^1)^2+(n_j^2)^2+(n_j^3)^2}$. The color scale of the arrows indicates $\langle \hat{S}^{z +}_j  \rangle= n_j^3$. The insets display the static spin structure factors $\mathcal{S}^{\perp}(\bm{\mathrm{q}})= \langle n^1_{\bm{\mathrm{q}}}n^1_{-\bm{\mathrm{q}}}+n^2_{\bm{\mathrm{q}}}n^2_{-\bm{\mathrm{q}}} \rangle$ and $\mathcal{S}^{zz}(\bm{\mathrm{q}})= \langle n^3_{\bm{\mathrm{q}}}n^3_{-\bm{\mathrm{q}}} \rangle$, with $\bm{\mathrm{n}}_{\bm{\mathrm{q}}}= \sum_{j} e^{\iu \bm{\mathrm{q}} \cdot \bm{\mathrm{r}}_j} \bm{\mathrm{n}}_j/L$. The $\mathbb{CP}^3$ skyrmion density distribution $\rho_{j k l}$ [see Eq.~\eqref{eq:totsky}] is indicated by the color of the triangular plaquettes.}
    \label{fig:double-q}
\end{figure*}

There are three different double-$ \bm{Q} $  orderings that are shown in Fig.~\ref{fig:phasediagram}, two of which (Fig.~\ref{fig:double-q}a,b) only appear in the non-perturbative regime. The third  (Fig.~\ref{fig:double-q}c) only appears in a very narrow region within the perturbative regime and has almost zero magnitude of dipolar components.   All three phases display striped structure in their $\mathbb{CP}^3$ skyrmion charge distribution. However, the 2Q-I phase exhibits stripes of chirality that oscillate smoothly from positive to negative values, whereas the 2Q-II and 2Q-III phases exhibit stripes that, e.g., start at large positive chirality, decrease in value to large negative chirality, and then have a stripe of zero chirality followed by a stripe of large positive chirality. 
Additionally, the 2Q-I and 2Q-II phases both contain rows of approximately $S^z=1$ triplet states in the locations where the chirality is changing smoothly from positive to negative, whereas the 2Q-III phase contains rows of approximately singlet states at these chirality transitions.

\subsection{Asymmetric Triple-Q Phases}

\begin{figure*}
    \centering
    \includegraphics[width=0.75\linewidth]{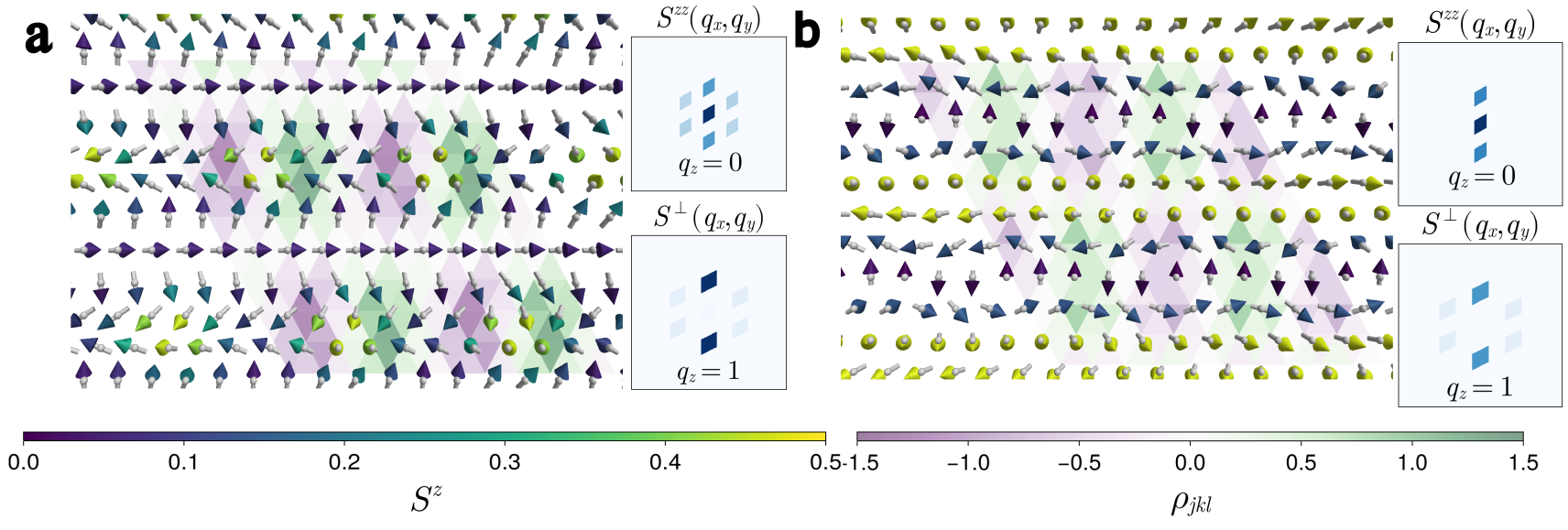}
    \captionsetup{justification=Justified}
    \caption{{\bf Real space distribution of the top layer dipolar sector of the two $\mathbb{CP}^3$ triple-{$\bm Q$} orderings.} {\bf a}. 3Q-I. {\bf b}. 3Q-II. The length of the arrow represents the magnitude of the dipole moment of the color field $|\langle \hat{\bm{\mathrm{S}}}_j ^{+}\rangle|=\sqrt{(n_j^1)^2+(n_j^2)^2+(n_j^3)^2}$. The color scale of the arrows indicates $\langle \hat{S}^{z +}_j  \rangle= n_j^3$. The insets display the static spin structure factors $\mathcal{S}^{\perp}(\bm{\mathrm{q}})= \langle n^1_{\bm{\mathrm{q}}}n^1_{-\bm{\mathrm{q}}}+n^2_{\bm{\mathrm{q}}}n^2_{-\bm{\mathrm{q}}} \rangle$ and $\mathcal{S}^{zz}(\bm{\mathrm{q}})= \langle n^3_{\bm{\mathrm{q}}}n^3_{-\bm{\mathrm{q}}} \rangle$, with $\bm{\mathrm{n}}_{\bm{\mathrm{q}}}= \sum_{j} e^{\iu \bm{\mathrm{q}} \cdot \bm{\mathrm{r}}_j} \bm{\mathrm{n}}_j/L$. The $\mathbb{CP}^3$ skyrmion density distribution $\rho_{j k l}$ [see Eq.~\eqref{eq:totsky}] is indicated by the color of the triangular plaquettes.}
    \label{fig:triple-Q}
\end{figure*}

Two distinct asymmetric triple-{$\bm Q$} spiral phases appear in Fig.~\ref{fig:phasediagram}, both characterized by a dominant contribution from one of the three ordering wave vectors $\bm{Q}_\nu$ ($\nu = 1, 2, 3$) [see Fig.~\ref{fig:triple-Q}]. In the 3Q-I phase [Fig.~\ref{fig:triple-Q}(a)], the longitudinal spin components exhibit nearly equal weight at all three $\bm{Q}_\nu$, while the transverse components are dominated by a single wave vector. In contrast, the 3Q-II phase [Fig.~\ref{fig:triple-Q}(b)] shows transverse components with one dominant wave vector and two others with small but finite amplitudes, whereas the longitudinal component has weight only at a single $\bm{Q}_\nu$. Both phases exhibit a checkerboard structure in their $\mathbb{CP}^3$ skyrmion charge density.

\section{$\mathfrak{su}(N)$ Algebra}
\label{app:SUN}

Here we give the derivations for the color field relations stated in Eq.~\eqref{eq:const} and ~\eqref{eq:Casimir} of the main paper but for the general case of generators of the $\mathfrak{su}(N)$ algebra.

There is a one-to-one correspondence between the operator field $\boldsymbol{\mathfrak{n}}_j$ and a coherent state $|\bm{\mathrm{Z}}_j\rangle$ up to some normalization constant $\kappa$ given by 

\begin{equation}
\boldsymbol{\mathfrak{n}}_j
=
\kappa\left(
|\bm{\mathrm{Z}}_j\rangle\langle\bm{\mathrm{Z}}_j|
-\frac{\mathbb{1}_N}{N}
\right).
\label{nzcorr}
\end{equation}
To determine $\kappa$, we take the Hilbert--Schmidt inner
product of Eq.~(F1) with $\hat{T}^\mu$. Using the
orthonormality condition
$\operatorname{Tr}(\hat{T}^\nu\hat{T}^\mu)=\delta_{\mu\nu}$
and the tracelessness of the generators gives $\kappa=1$.

To obtain the norm of the color field, we apply the definition of $n^\mu$ as well as Eq.~\eqref{nzcorr}:
\begin{eqnarray}
    n^\mu n^\mu &=& n^\mu \langle \bm{\mathrm{Z}}|{ \hat{T}}^\mu |\bm{\mathrm{Z}}\rangle 
    = \langle \bm{\mathrm{Z}}|\boldsymbol{\mathfrak{n}} |\bm{\mathrm{Z}}\rangle = \kappa \left( 1 - \frac{1}{N} \right).
\end{eqnarray}

Given the definition $d_{\mu \nu \eta} = \frac{1}{4} \text{Tr}({ \hat{T}}^\mu \{{\hat{T}}^\nu,{ \hat{T}}^\eta\})$, we derive Eq.~\ref{eq:const}:
\begin{eqnarray}
    d_{\mu \nu \eta} n^\nu n^\eta &=& \frac{1}{4}\text{Tr}({ \hat{T}}^\mu \{\boldsymbol{\mathfrak{n}},\boldsymbol{\mathfrak{n}}\}) =  \frac{1}{2} \text{Tr}\left({ \hat{T}}^\mu \left(\kappa \left(|\bm{\mathrm{Z}}\rangle \langle \bm{\mathrm{Z}}|- \mathbb{1}/N\right) \right)^2 \right) \nonumber \\
    &=& \frac{\kappa }{2} \left (\kappa - \frac{2}{N} \right) n^\mu. 
\end{eqnarray}
Using the above, we easily obtain the Casimir identity:
\begin{eqnarray}
    d_{\mu\nu\eta}n^\mu n^\nu n^\eta
    = \frac{\kappa^2}{2}\left (\kappa - \frac{2}{N} \right)\left(1- \frac{1}{N}\right).
\end{eqnarray}
For our particular case, where $\kappa = 1, N=4$ we obtain the correct coefficients used in the main text.

\section{Skyrmion Density in $\mathbb{CP}^3$}
\label{app:SkDensity}
The $\mathbb{CP}^3$ skyrmion charge is given by: 
\begin{equation}
C=\sum_{\triangle_{j k l}} \rho_{j k l}  = 
\frac{1}{2 \pi} \sum_{\triangle_{j k l}} \left(\gamma_{j l}+\gamma_{l k}+\gamma_{k j}\right),
\label{eq:totsky2}
\end{equation}
where $\triangle_{j k l}$ denotes each oriented triangular plaquette of nearest-neighbor sites $ j \to k \to l$
and 
$
\gamma_{k j}=\arg \left[\left\langle \bm{\mathrm{Z}}_{k} \mid \bm{\mathrm{Z}}_{j}\right\rangle\right]
$
is the Berry connection on the bond $j \to k$ and $\rho_{jkl}$ is the Berry phase on the triangle $jkl$~\cite{Hao23}.

We now show that the $\mathbb{CP}^3$ skyrmion flux is equivalent to the solid angle subtended on a suitably chosen Bloch sphere. (In the perturbative regime of our model, it is the one spanned by the singlet and $S^z = 1$ triplet states). First, consider three kets corresponding to the states of 3 adjacent dimers on the lattice: $|Z_1 \rangle, |Z_2 \rangle,$ and $|Z_3 \rangle$. Note that it is always possible to write $|Z_3 \rangle$ as a linear combination of the other two kets plus a third ket orthogonal to those two:
\begin{equation}
    |Z_3\rangle = \alpha |Z_1 \rangle + \beta |Z_2 \rangle + | Z_\perp{}\rangle,
\end{equation}
where $\langle Z_1 | Z_\perp{} \rangle = \langle Z_2 | Z_\perp{} \rangle = 0$. Furthermore, we can define a normalized state 
\[
|\tilde{Z}_3\rangle = \frac{P_{12}|Z_3\rangle}{\|P_{12}|Z_3\rangle\|},
\]
where \(P_{12}\) is the projector onto the subspace spanned by \(|Z_1\rangle\) and \(|Z_2\rangle\). By construction, \(P_{12}|Z_3\rangle = \alpha |Z_1\rangle + \beta |Z_2\rangle\).
Substituting this expression, we obtain:
\begin{eqnarray}
    \rho_{123}&=&\frac{1}{2\pi}\left(\arg[\langle Z_1 | Z_2 \rangle] + \arg[\langle Z_2 | Z_3 \rangle]  + \arg[\langle Z_3 | Z_1 \rangle]\right) \nonumber \\
    &=& \frac{1}{2\pi}\left(\arg[\langle Z_1 | Z_2 \rangle] + \arg[\langle Z_2 | \left( \alpha |Z_1 \rangle + \beta |Z_2 \rangle \right)] \right. \\  
    &+& \left. \arg[\left( \alpha^* \langle Z_1 | + \beta^* \langle Z_2 | \right) | Z_1 \rangle] \right). \nonumber
\end{eqnarray}

Now, suppose we pick the third ket to be $|\tilde{Z}_3 \rangle \in \mathbb{C}^2 $, where $\mathbb{C}^2$ is the subspace spanned by $|Z_1 \rangle$ and $|Z_2 \rangle$. Calculate the ``$\mathbb{CP}^1$ Skyrmion flux'' on the triangular plaquette $123$ and compare to the above result:
\begin{eqnarray}
    \rho_{12\tilde{3}}&=&\arg[\langle Z_1 | Z_2 \rangle] + \arg[\langle Z_2 | \tilde{Z}_3 \rangle]  + \arg[\langle \tilde{Z}_3 | Z_1 \rangle] \nonumber \\
    &=& \arg[\langle Z_1 | Z_2 \rangle] + \arg[\langle Z_2 | \left( \alpha |Z_1 \rangle + \beta |Z_2 \rangle  \right)] \nonumber \\ 
    &+& \arg[\left( \alpha^* \langle Z_1 | + \beta^* \langle Z_2 | \right)| Z_1 \rangle] \\
    &=& \rho_{123}. \nonumber
\end{eqnarray}
Thus, we have shown that the Berry phase $\rho_{123}$ on a given triangular plaquette 123 is equivalent to half of the solid angle subtended by the spherical triangle with vertices at \( |Z_1\rangle \), \( |Z_2\rangle \), and \( |\tilde{Z}_3\rangle \) on the Bloch sphere defined by the \(\mathbb{C}^2\) subspace spanned by \( |Z_1\rangle \) and \( |Z_2\rangle \).

\section{More SWT Results}
\label{app:DSSFextras}
\begin{figure*}
    \centering
    \includegraphics[width=1\linewidth]{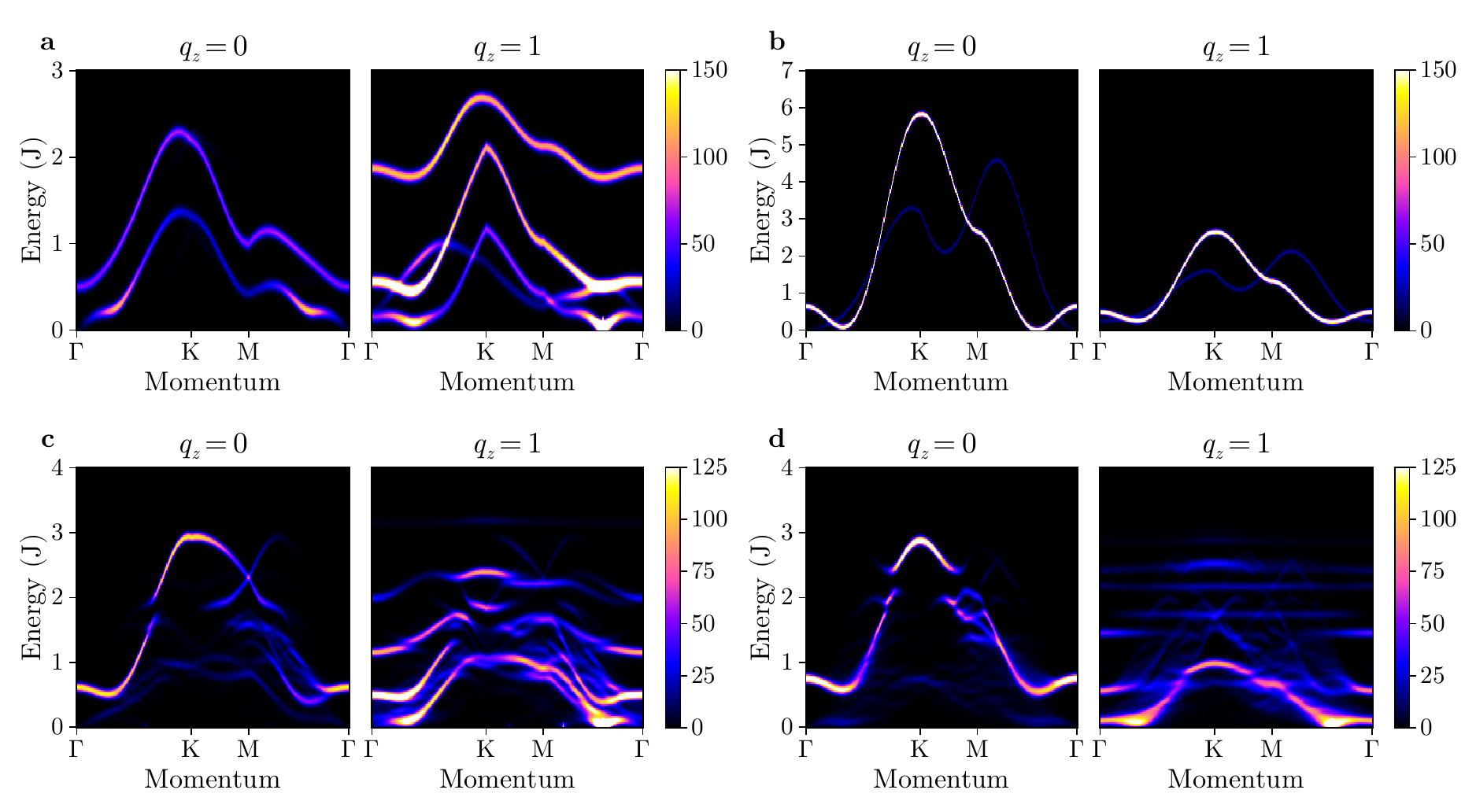}
    \captionsetup{justification=Justified}
    \caption{{\bf Magnetic excitations of the AFM-CS, FM-CS, 3Q-II, and SkX-II phases.} Each pair of panels shows the $q_z=0$ (left) and $q_z=1$ channels (right) of the dynamical spin structure factor (DSSF) calculated with linear spin wave theory from different phases of the phase diagram of Fig~\ref{fig:phasediagram}. {\bf a}. The DSSF of a system in the AFM-CS phase at the point $(\alpha,B)=(0.34,0.5)$. {\bf b}. The DSSF of a system in the FM-CS phase at the point $(\alpha,B)=(0.5,0.644)$. {\bf c}. The DSSF of a system in the 3$Q$-II phase $(\alpha,B)=(0.3,0.615)$.{\bf d}. The DSSF of a system in the SkX-II phase at the point $(\alpha,B)=(0.24,0.75)$. Calculations were performed using the \texttt{Sunny.jl} \cite{Dahlbom2025} package and include artificial Gaussian broadening with $\sigma=0.042$ J.}
    \label{fig:More_DSSF}
\end{figure*}

Fig.~\ref{fig:More_DSSF} includes DSSF calculations for the remaining single-$\bm{Q}$ phases as well as the SkX-II phase.

\begin{figure*}
    \centering
    \includegraphics[width=1\linewidth]{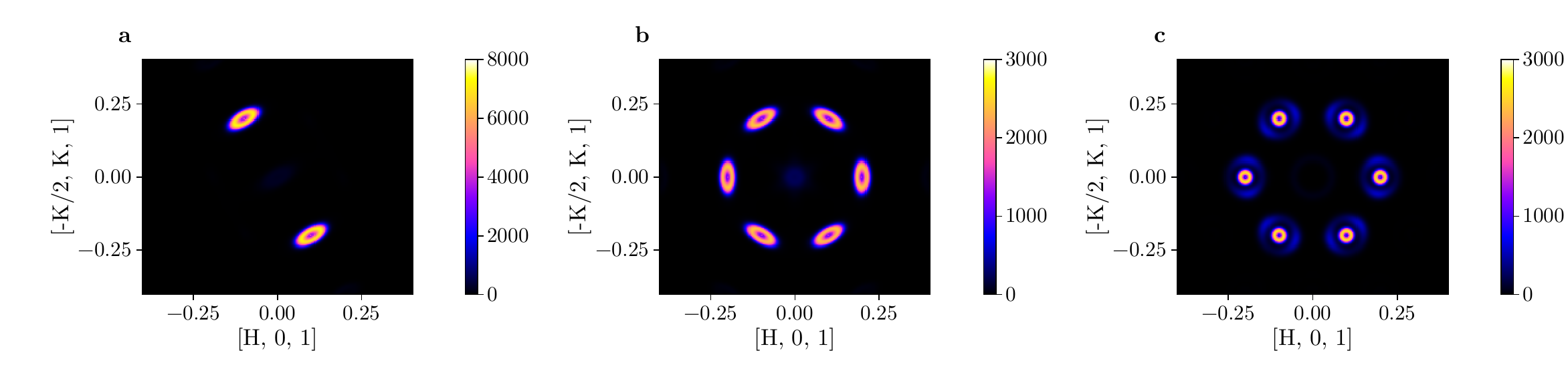}
    \captionsetup{justification=Justified}
    \caption{ {\bf Single-energy cuts of the YZ-Spiral and SkX-I phases at $\bm{\omega=0.012 J.}$} {\bf a}. Dynamical spin structure factor (DSSF) of a single domain of the YZ-Spiral phase at $(\alpha,B)=(0.1,0.8654)$. {\bf b}. DSSF of the same phase as {\bf a} but averaged over the three possible symmetry-equivalent domains. {\bf c}. DSSF for SkX-I phase at $(\alpha,B)=(0.14,0.7788)$. Note that the multi-domain single-Q ({\bf b}) and intrinsic triple-$Q$ ({\bf c}) phases can be distinguished by the cross sections at the ordering wave vectors, with the former being elliptical and the latter circular, as observed in \cite{Park25}. Calculations were performed using the \texttt{Sunny.jl} \cite{Dahlbom2025} package and include artificial Gaussian broadening with $\sigma=0.0042$ J.} 
    \label{fig:constant_energy_slices}
\end{figure*}

In Fig.~\ref{fig:constant_energy_slices}, we show constant-energy cuts of the DSSF near $\omega = 0$ for the YZ-spiral and SkX-I phases. These cuts highlight how single-$\bm{Q}$ and triple-$\bm{Q}$ magnetic orders can be distinguished through inelastic neutron scattering. In the single-$\bm{Q}$ phase, the transverse cross-section of each conical Goldstone mode appears elliptical, indicating a pronounced anisotropy in the mode velocities along the principal directions. In contrast, the triple-$\bm{Q}$ phase exhibits a nearly circular shape for its Goldstone mode, reflecting an emergent isotropy in the low-energy dispersion~\cite{Park25}.

\section{Numerical Methods}

The numerical procedure for obtaining the phase diagram in Fig.~\ref{fig:phasediagram} was identical to that described in \cite{Hao23}. A dense grid of points within the phase diagram was selected. For each point, a $5\times5$ cell of SU(4) coherent states was initialized. This choice of cell size is commensurate with the magnetic unit cell ($L=5$), which is fixed approximately by the parameter choice described below Eq.~\eqref{eq:alpha_def}. For each cell, the initial state was randomized and optimized using a gradient descent algorithm, resulting in a candidate ground state (local minimum) for each point of the phase diagram. This process was repeated using many different random initial conditions. The lowest-energy state discovered throughout the process was always retained, i.e., the best candidate at each point was updated only if the best state found in one iteration was better than the best state found in all previous iterations. After collecting ground state candidates for each point of the phase diagram, the procedure was extended. With each iteration, the optimization process was started with a random initial state and then successively with each of the optimal states found at the neighboring points of the phase diagram. This process was repeated until convergence was obtained. The numerical optimization within each iteration was performed using the \texttt{Sunny.jl} software package \cite{Dahlbom2025}, which enables direct implementation of the Hamiltonian Eq.~\eqref{HSU4} and calculates its gradient. \texttt{Sunny.jl} in turn relies on the conjugate gradient algorithm as implemented in the \texttt{Optim} package \cite{Mogensen2018}.

\begin{thebibliography}{10}
\expandafter\ifx\csname url\endcsname\relax
  \def\url#1{\texttt{#1}}\fi
\expandafter\ifx\csname urlprefix\endcsname\relax\def\urlprefix{URL }\fi
\providecommand{\bibinfo}[2]{#2}
\providecommand{\eprint}[2][]{\url{#2}}

\bibitem{Muhlbauer2009}
\bibinfo{author}{M\"{u}hlbauer, S.} \emph{et~al.}
\newblock \bibinfo{title}{Skyrmion lattice in a chiral magnet}.
\newblock \emph{\bibinfo{journal}{Science}} \textbf{\bibinfo{volume}{323}},
  \bibinfo{pages}{915} (\bibinfo{year}{2009}).
\newblock \urlprefix\url{http://www.sciencemag.org/content/323/5916/915}.

\bibitem{Yu2010a}
\bibinfo{author}{Yu, X.~Z.} \emph{et~al.}
\newblock \bibinfo{title}{Real-space observation of a two-dimensional skyrmion
  crystal}.
\newblock \emph{\bibinfo{journal}{Nature}} \textbf{\bibinfo{volume}{465}},
  \bibinfo{pages}{901} (\bibinfo{year}{2010}).
\newblock
  \urlprefix\url{http://www.nature.com/nature/journal/v465/n7300/full/nature09124.html}.

\bibitem{Yu2011}
\bibinfo{author}{Yu, X.~Z.} \emph{et~al.}
\newblock \bibinfo{title}{Near room-temperature formation of a skyrmion crystal
  in thin-films of the helimagnet {FeGe}}.
\newblock \emph{\bibinfo{journal}{Nature Materials}}
  \textbf{\bibinfo{volume}{10}}, \bibinfo{pages}{106} (\bibinfo{year}{2011}).
\newblock
  \urlprefix\url{http://www.nature.com/nmat/journal/v10/n2/full/nmat2916.html}.

\bibitem{Seki2012}
\bibinfo{author}{Seki, S.}, \bibinfo{author}{Yu, X.~Z.},
  \bibinfo{author}{Ishiwata, S.} \& \bibinfo{author}{Tokura, Y.}
\newblock \bibinfo{title}{Observation of skyrmions in a multiferroic material}.
\newblock \emph{\bibinfo{journal}{Science}} \textbf{\bibinfo{volume}{336}},
  \bibinfo{pages}{198} (\bibinfo{year}{2012}).
\newblock \urlprefix\url{http://www.sciencemag.org/content/336/6078/198}.

\bibitem{Adams2012}
\bibinfo{author}{Adams, T.} \emph{et~al.}
\newblock \bibinfo{title}{Long-wavelength helimagnetic order and skyrmion
  lattice phase in {$\mathrm{Cu_2OSeO_3}$}}.
\newblock \emph{\bibinfo{journal}{Phys. Rev. Lett.}}
  \textbf{\bibinfo{volume}{108}}, \bibinfo{pages}{237204}
  (\bibinfo{year}{2012}).
\newblock
  \urlprefix\url{http://link.aps.org/doi/10.1103/PhysRevLett.108.237204}.

\bibitem{Romming2013}
\bibinfo{author}{Romming, N.} \emph{et~al.}
\newblock \bibinfo{title}{Writing and deleting single magnetic skyrmions}.
\newblock \emph{\bibinfo{journal}{Science}} \textbf{\bibinfo{volume}{341}},
  \bibinfo{pages}{636--639} (\bibinfo{year}{2013}).
\newblock \urlprefix\url{http://www.sciencemag.org/content/341/6146/636}.

\bibitem{Bogolubskaya89}
\bibinfo{author}{Bogolubskaya, A.} \& \bibinfo{author}{Bogolubsky, I.}
\newblock \bibinfo{title}{Stationary topological solitons in the
  two-dimensional anisotropic heisenberg model with a skyrme term}.
\newblock \emph{\bibinfo{journal}{Physics Letters A}}
  \textbf{\bibinfo{volume}{136}}, \bibinfo{pages}{485--488}
  (\bibinfo{year}{1989}).

\bibitem{Bogolyubskaya89b}
\bibinfo{author}{Bogolyubskaya, A.~A.} \& \bibinfo{author}{Bogolyubsky, I.~L.}
\newblock \bibinfo{title}{{On Stationary Topological Solintons In
  Two-dimensional Anisotropic Heisenberg Model}}.
\newblock \emph{\bibinfo{journal}{Lett. Math. Phys.}}
  \textbf{\bibinfo{volume}{19}}, \bibinfo{pages}{171--177}
  (\bibinfo{year}{1990}).

\bibitem{Bogdanov1994}
\bibinfo{author}{Bogdanov, A.} \& \bibinfo{author}{Hubert, A.}
\newblock \bibinfo{title}{Thermodynamically stable magnetic vortex states in
  magnetic crystals}.
\newblock \emph{\bibinfo{journal}{Journal of Magnetism and Magnetic Materials}}
  \textbf{\bibinfo{volume}{138}}, \bibinfo{pages}{255 -- 269}
  (\bibinfo{year}{1994}).
\newblock
  \urlprefix\url{http://www.sciencedirect.com/science/article/pii/0304885394900469}.

\bibitem{Okubo12}
\bibinfo{author}{Okubo, T.}, \bibinfo{author}{Chung, S.} \&
  \bibinfo{author}{Kawamura, H.}
\newblock \bibinfo{title}{Multiple-$q$ states and the skyrmion lattice of the
  triangular-lattice heisenberg antiferromagnet under magnetic fields}.
\newblock \emph{\bibinfo{journal}{Phys. Rev. Lett.}}
  \textbf{\bibinfo{volume}{108}}, \bibinfo{pages}{017206}
  (\bibinfo{year}{2012}).
\newblock
  \urlprefix\url{https://link.aps.org/doi/10.1103/PhysRevLett.108.017206}.

\bibitem{Leonov2015}
\bibinfo{author}{Leonov, A.~O.} \& \bibinfo{author}{Mostovoy, M.}
\newblock \bibinfo{title}{Multiply periodic states and isolated skyrmions in an
  anisotropic frustrated magnet}.
\newblock \emph{\bibinfo{journal}{Nature Communications}}
  \textbf{\bibinfo{volume}{6}}, \bibinfo{pages}{8275} (\bibinfo{year}{2015}).
\newblock \urlprefix\url{https://doi.org/10.1038/ncomms9275}.

\bibitem{Lin2016_skyrmion}
\bibinfo{author}{Lin, S.-Z.} \& \bibinfo{author}{Hayami, S.}
\newblock \bibinfo{title}{Ginzburg-landau theory for skyrmions in
  inversion-symmetric magnets with competing interactions}.
\newblock \emph{\bibinfo{journal}{Phys. Rev. B}} \textbf{\bibinfo{volume}{93}},
  \bibinfo{pages}{064430} (\bibinfo{year}{2016}).
\newblock \urlprefix\url{https://link.aps.org/doi/10.1103/PhysRevB.93.064430}.

\bibitem{Hayami16}
\bibinfo{author}{Hayami, S.}, \bibinfo{author}{Lin, S.-Z.} \&
  \bibinfo{author}{Batista, C.~D.}
\newblock \bibinfo{title}{Bubble and skyrmion crystals in frustrated magnets
  with easy-axis anisotropy}.
\newblock \emph{\bibinfo{journal}{Phys. Rev. B}} \textbf{\bibinfo{volume}{93}},
  \bibinfo{pages}{184413} (\bibinfo{year}{2016}).
\newblock \urlprefix\url{https://link.aps.org/doi/10.1103/PhysRevB.93.184413}.

\bibitem{Batista16}
\bibinfo{author}{Batista, C.~D.}, \bibinfo{author}{Lin, S.-Z.},
  \bibinfo{author}{Hayami, S.} \& \bibinfo{author}{Kamiya, Y.}
\newblock \bibinfo{title}{Frustration and chiral orderings in correlated
  electron systems}.
\newblock \emph{\bibinfo{journal}{Reports on Progress in Physics}}
  \textbf{\bibinfo{volume}{79}}, \bibinfo{pages}{084504}
  (\bibinfo{year}{2016}).
\newblock
  \urlprefix\url{https://doi.org/10.1088%2F0034-4885%2F79%2F8%2F084504}.

\bibitem{Wang2020_RKKY}
\bibinfo{author}{Wang, Z.}, \bibinfo{author}{Su, Y.}, \bibinfo{author}{Lin,
  S.-Z.} \& \bibinfo{author}{Batista, C.~D.}
\newblock \bibinfo{title}{Skyrmion crystal from {{RKKY}} interaction mediated
  by {{2D}} electron gas}.
\newblock \emph{\bibinfo{journal}{Phys. Rev. Lett.}}
  \textbf{\bibinfo{volume}{124}}, \bibinfo{pages}{207201}
  (\bibinfo{year}{2020}).
\newblock
  \urlprefix\url{https://link.aps.org/doi/10.1103/PhysRevLett.124.207201}.

\bibitem{Hayami2021_review}
\bibinfo{author}{Hayami, S.} \& \bibinfo{author}{Motome, Y.}
\newblock \bibinfo{title}{Topological spin crystals by itinerant frustration}.
\newblock \emph{\bibinfo{journal}{J. Phys.: Condens. Matter}}
  \textbf{\bibinfo{volume}{33}}, \bibinfo{pages}{443001}
  (\bibinfo{year}{2021}).
\newblock \urlprefix\url{https://doi.org/10.1088/1361-648x/ac1a30}.

\bibitem{Yu2012}
\bibinfo{author}{Yu, X.~Z.} \emph{et~al.}
\newblock \bibinfo{title}{Skyrmion flow near room temperature in an ultralow
  current density}.
\newblock \emph{\bibinfo{journal}{Nature Communications}}
  \textbf{\bibinfo{volume}{3}}, \bibinfo{pages}{988} (\bibinfo{year}{2012}).
\newblock
  \urlprefix\url{http://www.nature.com/ncomms/journal/v3/n8/full/ncomms1990.html?WT.ec_id=NCOMMS-20120807}.

\bibitem{Yu2014_biskyrmion}
\bibinfo{author}{Yu, X.~Z.} \emph{et~al.}
\newblock \bibinfo{title}{Biskyrmion states and their current-driven motion in
  a layered manganite}.
\newblock \emph{\bibinfo{journal}{Nature Communications}}
  \textbf{\bibinfo{volume}{5}}, \bibinfo{pages}{3198} (\bibinfo{year}{2014}).
\newblock \urlprefix\url{https://doi.org/10.1038/ncomms4198}.

\bibitem{Mallik1998_paramana}
\bibinfo{author}{Mallik, R.}, \bibinfo{author}{Sampathkumaran, E.~V.},
  \bibinfo{author}{Paulose, P.~L.}, \bibinfo{author}{Sugawara, H.} \&
  \bibinfo{author}{Sato, H.}
\newblock \bibinfo{title}{Magnetic anomalies in gd$_2$pdsi$_3$}.
\newblock \emph{\bibinfo{journal}{Pramana - J. Phys.}}
  \textbf{\bibinfo{volume}{51}}, \bibinfo{pages}{505} (\bibinfo{year}{1998}).

\bibitem{Saha1999}
\bibinfo{author}{Saha, S.~R.} \emph{et~al.}
\newblock \bibinfo{title}{Magnetic anisotropy, first-order-like metamagnetic
  transitions, and large negative magnetoresistance in single-crystal
  \text{G}d$_{2}$\text{P}d\text{S}i$_{3}$}.
\newblock \emph{\bibinfo{journal}{Phys. Rev. B}} \textbf{\bibinfo{volume}{60}},
  \bibinfo{pages}{12162--12165} (\bibinfo{year}{1999}).
\newblock \urlprefix\url{https://link.aps.org/doi/10.1103/PhysRevB.60.12162}.

\bibitem{Kurumaji2019}
\bibinfo{author}{Kurumaji, T.} \emph{et~al.}
\newblock \bibinfo{title}{Skyrmion lattice with a giant topological hall effect
  in a frustrated triangular-lattice magnet}.
\newblock \emph{\bibinfo{journal}{Science}} \textbf{\bibinfo{volume}{365}},
  \bibinfo{pages}{914--918} (\bibinfo{year}{2019}).
\newblock \urlprefix\url{https://science.sciencemag.org/content/365/6456/914}.

\bibitem{Chandragiri_2016}
\bibinfo{author}{Chandragiri, V.}, \bibinfo{author}{Iyer, K.~K.} \&
  \bibinfo{author}{Sampathkumaran, E.~V.}
\newblock \bibinfo{title}{Magnetic behavior of
  \text{G}d$_3$\text{R}u$_4$\text{A}l$_{12}$, a layered compound with distorted
  kagom{\'{e}} net}.
\newblock \emph{\bibinfo{journal}{Journal of Physics: Condensed Matter}}
  \textbf{\bibinfo{volume}{28}}, \bibinfo{pages}{286002}
  (\bibinfo{year}{2016}).
\newblock
  \urlprefix\url{https://doi.org/10.1088%2F0953-8984%2F28%2F28%2F286002}.

\bibitem{Hirschberger2019}
\bibinfo{author}{Hirschberger, M.} \emph{et~al.}
\newblock \bibinfo{title}{Skyrmion phase and competing magnetic orders on a
  breathing kagom{\'e} lattice}.
\newblock \emph{\bibinfo{journal}{Nature Communications}}
  \textbf{\bibinfo{volume}{10}}, \bibinfo{pages}{5831} (\bibinfo{year}{2019}).
\newblock \urlprefix\url{https://doi.org/10.1038/s41467-019-13675-4}.

\bibitem{Hao21}
\bibinfo{author}{Zhang, H.} \& \bibinfo{author}{Batista, C.~D.}
\newblock \bibinfo{title}{Classical spin dynamics based on
  \text{S}\text{U}(\text{N}) coherent states}.
\newblock \emph{\bibinfo{journal}{Phys. Rev. B}}
  \textbf{\bibinfo{volume}{104}}, \bibinfo{pages}{104409}
  (\bibinfo{year}{2021}).
\newblock \urlprefix\url{https://link.aps.org/doi/10.1103/PhysRevB.104.104409}.

\bibitem{Hao23}
\bibinfo{author}{Zhang, H.}, \bibinfo{author}{Wang, Z.},
  \bibinfo{author}{Dahlbom, D.}, \bibinfo{author}{Barros, K.} \&
  \bibinfo{author}{Batista, C.~D.}
\newblock \bibinfo{title}{Cp2 skyrmions and skyrmion crystals in realistic
  quantum magnets}.
\newblock \emph{\bibinfo{journal}{Nature Communications}}
  \textbf{\bibinfo{volume}{14}}, \bibinfo{pages}{3626} (\bibinfo{year}{2023}).
\newblock \urlprefix\url{https://www.nature.com/articles/s41467-023-39232-8}.

\bibitem{akagi2021}
\bibinfo{author}{Akagi, Y.}, \bibinfo{author}{Amari, Y.},
  \bibinfo{author}{Sawado, N.} \& \bibinfo{author}{Shnir, Y.}
\newblock \bibinfo{title}{Isolated skyrmions in the {CP$^{2}$} nonlinear
  {$\sigma$}-model with a {Dzyaloshinskii}--{Moriya} type interaction}.
\newblock \emph{\bibinfo{journal}{Phys. Rev. D}}
  \textbf{\bibinfo{volume}{103}}, \bibinfo{pages}{065008}
  (\bibinfo{year}{2021}).

\bibitem{Zhang2024}
\bibinfo{author}{Zhang, H.} \& \bibinfo{author}{Lin, S.-Z.}
\newblock \bibinfo{title}{Multipolar skyrmion crystals in non-kramers doublet
  systems}.
\newblock \emph{\bibinfo{journal}{Physical Review Letters}}
  \textbf{\bibinfo{volume}{133}}, \bibinfo{pages}{196702}
  (\bibinfo{year}{2024}).

\bibitem{LinHayami2016}
\bibinfo{author}{Lin, S.-Z.} \& \bibinfo{author}{Hayami, S.}
\newblock \bibinfo{title}{Ginzburg--landau theory for skyrmions in
  inversion-symmetric magnets with competing interactions}.
\newblock \emph{\bibinfo{journal}{Physical Review B}}
  \textbf{\bibinfo{volume}{93}}, \bibinfo{pages}{064430}
  (\bibinfo{year}{2016}).

\bibitem{Perelomov1972}
\bibinfo{author}{Perelomov, A.~M.}
\newblock \bibinfo{title}{Coherent states for arbitrary lie groups}.
\newblock \emph{\bibinfo{journal}{Communications in Mathematical Physics}}
  \textbf{\bibinfo{volume}{26}}, \bibinfo{pages}{222--236}
  (\bibinfo{year}{1972}).

\bibitem{David2022}
\bibinfo{author}{Dahlbom, D.} \emph{et~al.}
\newblock \bibinfo{title}{Geometric integration of classical spin dynamics via
  a mean-field schr\"odinger equation}.
\newblock \emph{\bibinfo{journal}{Phys. Rev. B}}
  \textbf{\bibinfo{volume}{106}}, \bibinfo{pages}{054423}
  (\bibinfo{year}{2022}).
\newblock \urlprefix\url{https://link.aps.org/doi/10.1103/PhysRevB.106.054423}.

\bibitem{David2022b}
\bibinfo{author}{Dahlbom, D.}, \bibinfo{author}{Miles, C.},
  \bibinfo{author}{Zhang, H.}, \bibinfo{author}{Batista, C.~D.} \&
  \bibinfo{author}{Barros, K.}
\newblock \bibinfo{title}{Langevin dynamics of generalized spins as su($n$)
  coherent states} (\bibinfo{year}{2022}).
\newblock \urlprefix\url{https://arxiv.org/abs/2209.01265}.

\bibitem{David2023}
\bibinfo{author}{Dahlbom, D.} \emph{et~al.}
\newblock \bibinfo{title}{Renormalized classical theory of quantum magnets}
  (\bibinfo{year}{2023}).
\newblock \urlprefix\url{https://arxiv.org/abs/2304.03874}.

\bibitem{Gnutzmann98}
\bibinfo{author}{Gnutzmann, S.} \& \bibinfo{author}{Kus, M.}
\newblock \bibinfo{title}{Coherent states and the classical limit on
  irreducible representations}.
\newblock \emph{\bibinfo{journal}{Journal of Physics A: Mathematical and
  General}} \textbf{\bibinfo{volume}{31}}, \bibinfo{pages}{9871--9896}
  (\bibinfo{year}{1998}).
\newblock \urlprefix\url{https://doi.org/10.1088/0305-4470/31/49/011}.

\bibitem{Papanicolaou1988}
\bibinfo{author}{Papanicolaou, N.}
\newblock \bibinfo{title}{Unusual phases in quantum spin-1 systems}.
\newblock \emph{\bibinfo{journal}{Nuclear Physics B}}
  \textbf{\bibinfo{volume}{305}}, \bibinfo{pages}{367--395}
  (\bibinfo{year}{1988}).
\newblock
  \urlprefix\url{https://www.sciencedirect.com/science/article/pii/0550321388900739}.

\bibitem{Batista04}
\bibinfo{author}{Batista, C.~D.} \& \bibinfo{author}{Ortiz, G.}
\newblock \bibinfo{title}{Algebraic approach to interacting quantum systems}.
\newblock \emph{\bibinfo{journal}{Advances in Physics}}
  \textbf{\bibinfo{volume}{53}}, \bibinfo{pages}{1--82} (\bibinfo{year}{2004}).
\newblock \urlprefix\url{https://doi.org/10.1080/00018730310001642086}.
\newblock \eprint{https://doi.org/10.1080/00018730310001642086}.

\bibitem{Zapf06}
\bibinfo{author}{Zapf, V.~S.} \emph{et~al.}
\newblock \bibinfo{title}{Bose-\text{E}instein condensation of $s=1$ nickel
  spin degrees of freedom in
  \text{N}i\text{C}l$_{2}\mathrm{\text{\ensuremath{-}}}4\mathrm{SC}$(\text{N}\text{H}$_{2}{)}_{2}$}.
\newblock \emph{\bibinfo{journal}{Phys. Rev. Lett.}}
  \textbf{\bibinfo{volume}{96}}, \bibinfo{pages}{077204}
  (\bibinfo{year}{2006}).
\newblock
  \urlprefix\url{https://link.aps.org/doi/10.1103/PhysRevLett.96.077204}.

\bibitem{Lauchli2006}
\bibinfo{author}{L{\"a}uchli, A.}, \bibinfo{author}{Mila, F.} \&
  \bibinfo{author}{Penc, K.}
\newblock \bibinfo{title}{Quadrupolar {{Phases}} of the ${{S}}=1$
  {{Bilinear-Biquadratic Heisenberg Model}} on the {{Triangular Lattice}}}.
\newblock \emph{\bibinfo{journal}{Phys. Rev. Lett.}}
  \textbf{\bibinfo{volume}{97}}, \bibinfo{pages}{087205}
  (\bibinfo{year}{2006}).
\newblock
  \urlprefix\url{https://link.aps.org/doi/10.1103/PhysRevLett.97.087205}.

\bibitem{Muniz14}
\bibinfo{author}{Muniz, R.~A.}, \bibinfo{author}{Kato, Y.} \&
  \bibinfo{author}{Batista, C.~D.}
\newblock \bibinfo{title}{Generalized spin-wave theory: {{Application}} to the
  bilinear\textendash biquadratic model}.
\newblock \emph{\bibinfo{journal}{Prog. Theor. Exp. Phys.}}
  \textbf{\bibinfo{volume}{2014}}, \bibinfo{pages}{083I01}
  (\bibinfo{year}{2014}).
\newblock \urlprefix\url{https://doi.org/10.1093/ptep/ptu109}.

\bibitem{Galkina14}
\bibinfo{author}{Galkina, E.~G.}, \bibinfo{author}{Ivanov, B.~A.} \&
  \bibinfo{author}{Butrim, V.~I.}
\newblock \bibinfo{title}{Longitudinal spin dynamics in nickel fluorosilicate}.
\newblock \emph{\bibinfo{journal}{Low Temperature Physics}}
  \textbf{\bibinfo{volume}{40}}, \bibinfo{pages}{635--640}
  (\bibinfo{year}{2014}).
\newblock \urlprefix\url{https://doi.org/10.1063/1.4890989}.
\newblock \eprint{https://doi.org/10.1063/1.4890989}.

\bibitem{David2024}
\bibinfo{author}{Dahlbom, D.}, \bibinfo{author}{Thomas, J.},
  \bibinfo{author}{Johnston, S.}, \bibinfo{author}{Barros, K.} \&
  \bibinfo{author}{Batista, C.~D.}
\newblock \bibinfo{title}{Classical dynamics of the antiferromagnetic
  heisenberg s = 1/2 spin ladder}.
\newblock \emph{\bibinfo{journal}{Phys. Rev. B}}
  \textbf{\bibinfo{volume}{110}}, \bibinfo{pages}{104403}
  (\bibinfo{year}{2024}).
\newblock
  \urlprefix\url{https://journals.aps.org/prb/abstract/10.1103/PhysRevB.110.104403}.

\bibitem{Dahlbom2025}
\bibinfo{author}{Dahlbom, D.} \emph{et~al.}
\newblock \bibinfo{title}{Sunny.jl: A julia package for spin dynamics}
  (\bibinfo{year}{2025}).
\newblock \urlprefix\url{https://arxiv.org/abs/2501.13095}.
\newblock \eprint{2501.13095}.

\bibitem{Park25}
\bibinfo{author}{Park, P.} \emph{et~al.}
\newblock \bibinfo{title}{Contrasting dynamical properties of single-q and
  triple-q magnetic orderings in a triangular lattice antiferromagnet}
  (\bibinfo{year}{2025}).
\newblock \urlprefix\url{https://arxiv.org/abs/2410.02180}.
\newblock \eprint{2410.02180}.

\bibitem{Feng2020}
\bibinfo{author}{Feng, W.} \emph{et~al.}
\newblock \bibinfo{title}{Topological magneto-optical effects and their
  quantization in noncoplanar antiferromagnets}.
\newblock \emph{\bibinfo{journal}{Nature Communications}}
  \textbf{\bibinfo{volume}{11}}, \bibinfo{pages}{118} (\bibinfo{year}{2020}).

\bibitem{Dahlbom2022b}
\bibinfo{author}{Dahlbom, D.}, \bibinfo{author}{Miles, C.},
  \bibinfo{author}{Zhang, H.}, \bibinfo{author}{Batista, C.~D.} \&
  \bibinfo{author}{Barros, K.}
\newblock \bibinfo{title}{Langevin dynamics of generalized spins as su($n$)
  coherent states}.
\newblock \emph{\bibinfo{journal}{Physical Review B}}
  \textbf{\bibinfo{volume}{106}}, \bibinfo{pages}{235154}
  (\bibinfo{year}{2022}).
\newblock \urlprefix\url{https://link.aps.org/doi/10.1103/PhysRevB.106.235154}.

\bibitem{stone2008}
\bibinfo{author}{Stone, M.~B.} \emph{et~al.}
\newblock \bibinfo{title}{Singlet-triplet dispersion reveals additional
  frustration in the triangular-lattice dimer compound ba3mn2o8}.
\newblock \emph{\bibinfo{journal}{Physical Review Letters}}
  \textbf{\bibinfo{volume}{100}}, \bibinfo{pages}{237201}
  (\bibinfo{year}{2008}).

\bibitem{samulon2009}
\bibinfo{author}{Samulon, E.~C.} \emph{et~al.}
\newblock \bibinfo{title}{Asymmetric quintuplet condensation in the frustrated
  s=1 spin dimer compound ba3mn2o8}.
\newblock \emph{\bibinfo{journal}{Phys. Rev. Lett.}}
  \textbf{\bibinfo{volume}{103}}, \bibinfo{pages}{047202}
  (\bibinfo{year}{2009}).

\bibitem{samulon2010}
\bibinfo{author}{Samulon, E.~C.} \emph{et~al.}
\newblock \bibinfo{title}{Anisotropic phase diagram of the frustrated spin
  dimer compound ba3mn2o8}.
\newblock \emph{\bibinfo{journal}{Phys. Rev. B}} \textbf{\bibinfo{volume}{81}},
  \bibinfo{pages}{104421} (\bibinfo{year}{2010}).

\bibitem{Lee2014}
\bibinfo{author}{Lee, M.} \emph{et~al.}
\newblock \bibinfo{title}{Series of phase transitions and multiferroicity in
  the quasi-two-dimensional spin-$\frac{1}{2}$ triangular-lattice
  antiferromagnet ${\mathrm{ba}}_{3}{\mathrm{conb}}_{2}{\mathrm{o}}_{9}$}.
\newblock \emph{\bibinfo{journal}{Phys. Rev. B}} \textbf{\bibinfo{volume}{89}},
  \bibinfo{pages}{104420} (\bibinfo{year}{2014}).
\newblock \urlprefix\url{https://link.aps.org/doi/10.1103/PhysRevB.89.104420}.

\bibitem{Yokota2014}
\bibinfo{author}{Yokota, K.}, \bibinfo{author}{Kurita, N.} \&
  \bibinfo{author}{Tanaka, H.}
\newblock \bibinfo{title}{Magnetic phase diagram of the $s=\frac{1}{2}$
  triangular-lattice heisenberg antiferromagnet
  ${\mathrm{ba}}_{3}\mathrm{Co}{\mathrm{nb}}_{2}{\mathrm{o}}_{9}$}.
\newblock \emph{\bibinfo{journal}{Phys. Rev. B}} \textbf{\bibinfo{volume}{90}},
  \bibinfo{pages}{014403} (\bibinfo{year}{2014}).
\newblock \urlprefix\url{https://link.aps.org/doi/10.1103/PhysRevB.90.014403}.

\bibitem{Johnson1979}
\bibinfo{author}{Johnson, P.}, \bibinfo{author}{Rayne, J.} \&
  \bibinfo{author}{Friedberg, S.}
\newblock \bibinfo{title}{Magnetic properties of csnicl3 and rbnicl3}.
\newblock \emph{\bibinfo{journal}{Journal of Applied Physics}}
  \textbf{\bibinfo{volume}{50}}, \bibinfo{pages}{1853--1855}
  (\bibinfo{year}{1979}).

\bibitem{Note1}
\bibinfo{note}{This manuscript has been authored by UT-Battelle, LLC, under
  contract DE-AC05-00OR22725 with the US Department of Energy (DOE). The US
  government retains and the publisher, by accepting the article for
  publication, acknowledges that the US government retains a nonexclusive,
  paid-up, irrevocable, worldwide license to publish or reproduce the published
  form of this manuscript, or allow others to do so, for US government
  purposes. DOE will provide public access to these results of federally
  sponsored research in accordance with the DOE Public Access Plan
  (https://www.energy.gov/doe-public-access-plan)}.

\bibitem{Mogensen2018}
\bibinfo{author}{Mogensen, P.~K.} \& \bibinfo{author}{Riseth, A.~N.}
\newblock \bibinfo{title}{Optim: A mathematical optimization package for
  julia}.
\newblock \emph{\bibinfo{journal}{Journal of Open Source Software}}
  \textbf{\bibinfo{volume}{3}}, \bibinfo{pages}{615} (\bibinfo{year}{2018}).
\newblock \urlprefix\url{https://doi.org/10.21105/joss.00615}.

\end{thebibliography}
\end{document}